\documentclass[dvipsnames,twocolumn]{aastex63}
\PassOptionsToPackage{usenames,dvipsnames}{xcolor}

\usepackage{tabularx}
\usepackage{amsmath}  
\usepackage{amssymb}  
\usepackage[T1]{fontenc}

\submitjournal{ApJ}
\accepted{January 4, 2021}
\shorttitle{Strongly Lensed SN Cosmology with the Roman Space Telescope}
\shortauthors{Pierel et al.}

\newif{\ifchangetext}
\changetextfalse

\InputIfFileExists{\jobname-options}

\ifchangetext
  
  \newcommand{\changenote}[1]{\textcolor{blue}{ \bf #1}}x
\else
  
  \newcommand{\changenote}[1]{}
\fi

\hypersetup{
    colorlinks=True,
    citecolor=black,
    linkcolor=black,
    filecolor=magenta,      
    urlcolor=cyan,
}

\def\xone{\ensuremath{x_{1}}}
\def\C{\ensuremath{\mathcal{C}}}

\def\TC{\ensuremath{\mathcal{T_C}}}
\def\TL{\ensuremath{\mathcal{T_L}}}
\def\dTC{\ensuremath{\delta\mathcal{T_C}}}
\def\dTL{\ensuremath{\delta\mathcal{T_L}}}

\defcitealias{hounsell_simulations_2018}{H18}

\defcitealias{oguri_gravitationally_2010}{OM10}

\defcitealias{pierel_turning_2019}{PR19}

\begin{document}

\title{\Large Projected Cosmological Constraints from Strongly Lensed Supernovae with the \textit{Roman Space Telescope}}

\correspondingauthor{J.~D.~R.~Pierel}
\email{jr23@email.sc.edu}
\author{J.~D.~R.~Pierel}

\affil{Department of Physics and Astronomy, University of South Carolina, 712 Main St., Columbia, SC 29208, USA}
\author{S.~Rodney}

\affil{Department of Physics and Astronomy, University of South Carolina, 712 Main St., Columbia, SC 29208, USA}
\author{G.~Vernardos}

\affil{Institute of Astrophysics, Foundation for Research and Technology - Hellas (FORTH), GR-70013,Heraklion, Greece}
\author{M.~Oguri}

\affil{Research Center for the Early Universe, University of Tokyo, Tokyo 113-0033, Japan}
\affil{Department of Physics, University of Tokyo, Tokyo 113-0033, Japan}
\affil{Kavli Institute for the Physics and Mathematics of the Universe (Kavli IPMU, WPI), University of Tokyo, Chiba 277-8582, Japan}
\author{R.~Kessler}

\affil{Department of Astronomy and Astrophysics, University of Chicago, Chicago, IL 60637, USA}
\affil{Kavli Institute for Cosmological Physics, University of Chicago, Chicago, IL 60637, USA}
\author{T.~Anguita}

\affil{Departamento de Ciencias Fisicas, Universidad Andres Bello Fernandez Concha 700, 7591538 Las Condes, Santiago, Chile}
\affil{Millennium Institute of Astrophysics, Monseñor Nuncio Sotero Sanz 100, Oficina 104, 7500011 Providencia, Santiago, Chile}

\begin{abstract}
One of the primary mission objectives for the \textit{Roman Space Telescope} is to investigate the nature of dark energy with a variety of methods. 
Observations of Type Ia supernovae (SNIa) will be one of the principal anchors of the \textit{Roman} cosmology program, through traditional luminosity distance measurements. 
This SNIa cosmology program can provide another valuable cosmological probe, without altering the mission strategy: time delay cosmography with gravitationally lensed SN. In this work, we forecast lensed SN cosmology constraints with the \textit{Roman Space Telescope}, while providing useful tools for future work. 
Using anticipated characteristics of the \textit{Roman} SNIa survey, we have constructed mock catalogs of expected resolved lensing systems, as well as strongly lensed Type Ia and core-collapse (CC) SN light curves, including microlensing effects. We predict \textit{Roman} will find $\sim11$ lensed SNIa and $\sim20$ CCSN, dependent on the survey strategy. Next, we estimate the time delay precision obtainable with \textit{Roman} (Ia: $\sim2 $ days, CC: $\sim3$ days), and use a Fisher matrix analysis to derive projected constraints on $H_0,\Omega_m$, and the dark energy Equation of State (EOS), $w$, for each SNIa survey strategy.  A strategy optimized for high-redshift SNIa discovery is preferred when considering the constraints possible from both SNIa and lensed SN cosmology, also delivering $\sim1.5$ times more lensed SN than other proposed survey strategies. \end{abstract}

\section{Introduction}
\label{sec:intro}
The theory necessary for using gravitationally lensed SN resolved into multiple images (lensed SN) as cosmological probes has been in place for decades \citep{refsdal_possibility_1964}. However, the first lensed SN were only discovered in the past 6 years \citep{quimby_detection_2014,kelly_multiple_2015,goobar_iptf16geu:_2017}.  These rare events manifest when the light from a stellar explosion propagating along different paths is focused by a lensing potential (a galaxy or galaxy cluster), forming multiple images of the SN on the sky. Depending on the relative geometrical and gravitational potential differences of each path, the SN images appear delayed by hours to months (for galaxy-scale lenses) or years (for cluster-scale lenses). These time delays can be used to measure a ratio of angular diameter distances which in turn constrains the Hubble-LeMa\^itre Constant, $H_0$, and the  dark energy Equation of State (EOS) $w$ \citep{linder_lensing_2011,treu_time_2016}.

 Time delay cosmography has been employed  for decades using multiply-imaged quasars \citep[e.g.,][]{vuissoz_cosmograil_2008,suyu_dissecting_2010,tewes_cosmograil_2013,bonvin_h0licow_2017,birrer_h0licow_2019,bonvin_cosmograil_2018,bonvin_cosmograil_2019}.  Lensed SN have several unique characteristics relative to lensed quasars that are advantageous for time delay measurement \citep[see][for a review]{oguri_strong_2019}. For example, most SN exhibit a single luminosity peak and evolve over short time-scales \citep{woosley_type_2007,sanders_toward_2015}, some have a standardizable absolute brightness \citep{phillips_absolute_1993,hamuy_type_2002,poznanski_improved_2009,kasen_type_2009}, and there are well-calibrated SN light curve models for fitting purposes \citep{jha_improved_2007,guy_salt2:_2007,burns_carnegie_2011}. In addition to these advantages in measuring time delays, intrinsic brightness can be inferred for some lensed SN independent of lensing \citep[though only in cases where millilensing and microlensing are not extreme, see][]{goobar_iptf16geu:_2017,foxley-marrable_impact_2018,dhawan_magnification_2019}. This could provide additional leverage for lens modeling by helping to limit the uncertainty caused by the mass-sheet degeneracy \citep{falco_model_1985,kolatt_gravitational_1998,holz_seeing_2001,oguri_gravitational_2003,rodney_illuminating_2015,xu_lens_2016}. Every lensed SN also eventually fades away and leaves the scene, which is advantageous for modeling the lensing potential, especially in galaxy-scale lensing systems.

Although the current sample of lensed SN is at present only two, the \textit{Roman Space Telescope} (previously WFIRST) is expected to detect dozens of lensed SN after launching in the mid-2020's \citep{oguri_gravitationally_2010}. \textit{Roman} will be extremely valuable for lensed SN time delay cosmography because of its unique combination of large field of view, 
sharp point-spread function (PSF), and wide redshift range. Lensed SN discovered by \textit{Roman} will already have images of sufficient quality to build a lens model, which usually requires a separate follow-up campaign \citep[e.g.,][]{bonvin_h0licow_2017,dhawan_magnification_2019}. 

The goal of this work is  to understand the value of various SN survey strategies and what will be achievable with these lensed SN. We create realistic forecasts of \textit{Roman Space Telescope} time delay cosmography using simulated lensed SN, including and analyzing microlensing effects that can perturb SN light curves by as much as half a magnitude or more \citep{dobler_microlensing_2006,goldstein_precise_2018}. This requires estimates of the number of lensed SN that will be observed, and the time delay precision obtainable for each system. There have been no investigations to date that use \textit{Roman}'s instrument and survey properties to estimate the lensed SN yield, or to create detailed simulations of macro- and micro-lensed SN light curves that accurately mimic those we expect to observe during the SNIa survey. Here we produce both for the first time, making each resource publicly available, and propagate the results through to projections for cosmology.

We begin in Section \ref{sec:NGRST} by predicting lensed SN yields for various \textit{Roman Space Telescope} SN survey strategies, and producing mock lens catalogs for use in future work. In Sections \ref{sec:sims}-\ref{sec:delays} we simulate a large number of SN light curves using current parameters for \textit{Roman} instruments and surveys. Next, we fit these light curves to estimate the time delay precision and accuracy for these lensed SN. Finally, we propagate the results from the sections listed above through to cosmological constraints in Section \ref{sec:cosmo}, completing the picture of SN time delay cosmography with the \textit{Roman Space Telescope}. A discussion of the results follows in Section \ref{sec:discussion}.

\section{Simulating Lens Catalogs for the Roman Space Telescope}
\label{sec:NGRST}
This section describes our simulated lens catalogs used to determine the expected lensed SN yield for various proposed surveys for the \textit{Roman Space Telescope} mission. We provide a brief description of the surveys evaluated in this work, and describe the methodology used to create libraries of mock lenses and calculated lensed SN yields for each strategy.

\subsection{\textit{Roman Space Telescope} Supernova Surveys}
\label{sub:sn_surveys}

The survey strategies investigated in this work are based on the simulations of \citet[][hereafter \citetalias{hounsell_simulations_2018}]{hounsell_simulations_2018}, who sought to make realistic simulations of the \textit{Roman Space Telescope} SN survey, based upon the hardware specifications at the time and a variety of survey strategies. The \citetalias{hounsell_simulations_2018} simulations do not include images, but they include forecasted image properties such as detection efficiency vs. magnitude, sky noise, readout noise, and point spread function (PSF). There were 11 strategies investigated including 4 from the original Science Definition Team (SDT) final report \citep{spergel_wide_2015}, and 4 variations on each of 3 baseline strategies. 
 Since the publication of \citetalias{hounsell_simulations_2018}, the integral field channel (IFC) has been removed from the design of \textit{Roman}. 
 Therefore, for this work we  use only the 3 baseline ``Imaging'' strategies that assumed no time allocated to the IFC.   We also note that some \textit{Roman} SNIa survey designs under consideration may allocate a significant fraction of the observing time to slitless spectroscopy with the prism. Consideration of such designs would require analysis of the value of prism spectroscopy for SN time delay measurement.  We defer such explorations to future work.

 The three imaging strategies we consider, labeled Lowz, Allz, and Highz, were designed to emphasize SNIa discovery in different redshift ranges (as their names indicate).  Each strategy is constructed as a series of ``tiers'' labeled as Shallow, Medium, and Deep (Table \ref{tab:rst_survey}). Each tier was simulated over a redshift range of $0.01-2.99$, using RZYJ filters for the Shallow and Medium tiers, and YJHF for the Deep tier (see \citetalias{hounsell_simulations_2018} Figure 1 for filter transmission details). 
 \begin{figure*}[ht!]
\centering
\includegraphics[ width=0.49\textwidth]{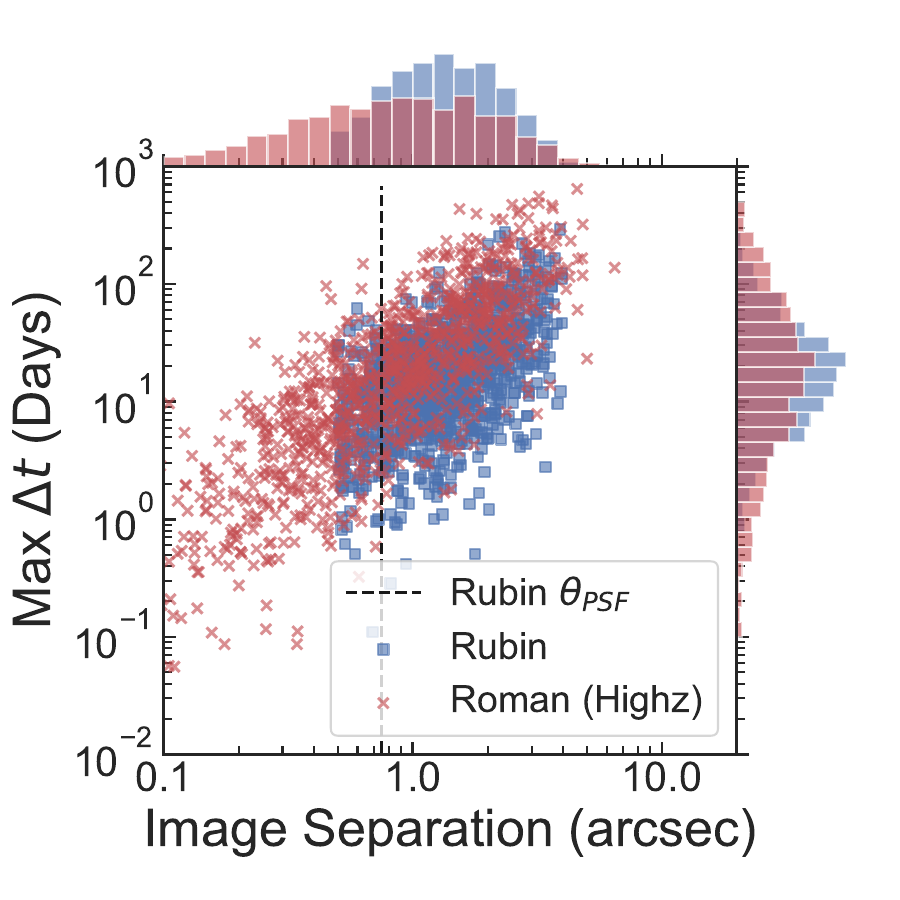}
\includegraphics[ width=0.49\textwidth]{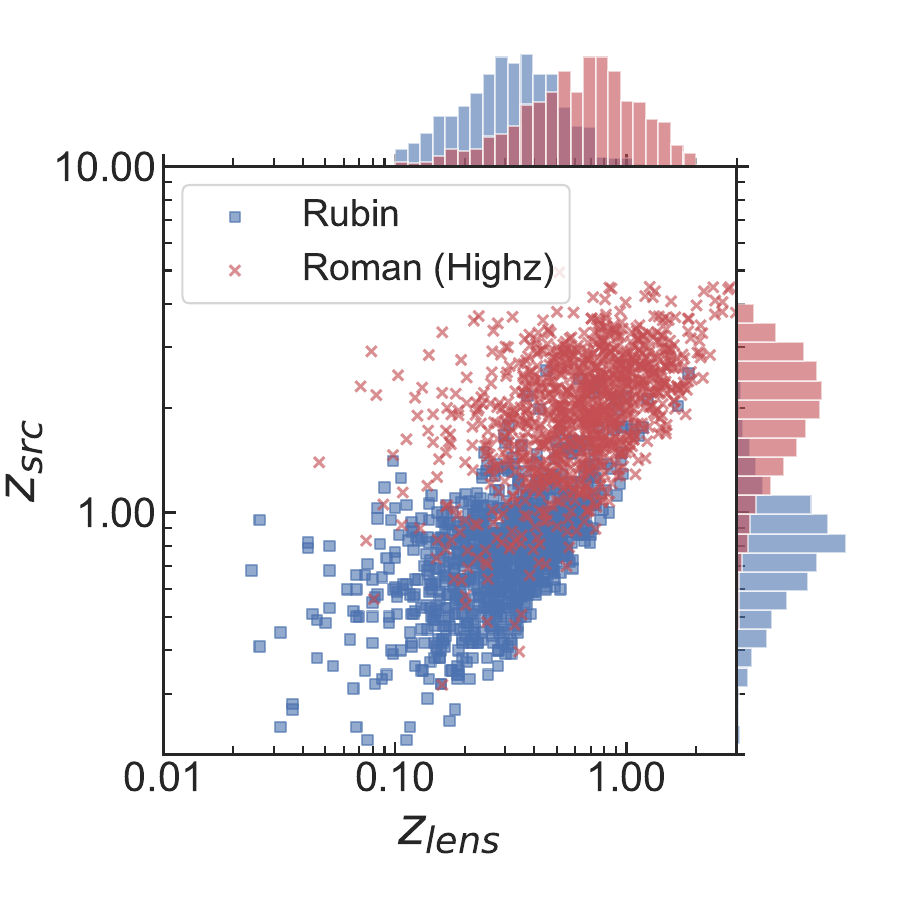}
\vspace{-10pt}
\caption{\label{fig:survey_redshifts} Comparison of properties between the \citetalias{oguri_gravitationally_2010} \textit{Vera Rubin} LSST mock lens catalog and the Highz lens catalog created in this work for the \textit{Roman Space Telescope}. We use 1000 random lenses from each catalog for comparison, though the actual projected yields for \textit{Rubin} are much higher than for \textit{Roman} \citep{goldstein_rates_2019,huber_strongly_2019}. In both figures, \textit{Rubin} data are shown in blue and \textit{Roman} data are shown in red. (Left) The maximum time delay for each mock vs. the maximum separation between multiple images of the source in arcseconds. The grey dashed line represents the likely point spread function (PSF) full width at half maximum (FWHM) for \textit{Rubin}; images are ``resolved'' when their separation is $\geq2/3~\theta_{PSF}$. (Right) The source redshift vs. the lens redshift for each lensing system. The \textit{Rubin} mock lens catalog uses a 23.3 mag (i-Band) detection threshold, while we use 26.2 mag (H-Band) for \textit{Roman} (see Table \ref{tab:rst_survey}).}
\end{figure*}
 An important  modification introduced in this work for simulating these strategies is that we extend the redshift range to $z=4$, due to the possibility of detecting such high-redshift SN if they are bright and significantly magnified \citep[see Section \ref{sub:lens}]{oguri_gravitationally_2010}. The estimated SNIa yields found by \citetalias{hounsell_simulations_2018} for each survey strategy can be found in Table \ref{tab:rst_survey}.

In addition to the \citetalias{hounsell_simulations_2018} survey strategies described above, we create mock lens catalogs (and predict yields) for two more surveys. The first is the High Latitude Survey (HLS), which will enable \textit{Roman Space Telescope} weak lensing shape measurements \citep{spergel_wide_2015}. The HLS will image 2000 $\rm{deg}^2$ in at least two passes, separated in time by at least several months.  This will enable a single-epoch transient search, though delivery of difference images is not an expected part of HLS operations. Because of its wide survey area and deep imaging, the HLS is  capable of detecting intrinsically rare transient events, such as lensed SN.  Any lensed SN discovered in the HLS would require a separate follow-up campaign. We therefore include the HLS for calculating lensed SN yields, but leave the investigation of a follow-up strategy to future work.

The second additional strategy we investigate is optimized for lensed SN discovery, as we find that depth is the most important factor in achieving a higher lensed SN yield. We name this survey the ``UltraDeep'' strategy, and use it as an upper limit of what the \textit{Roman Space Telescope} could be capable of discovering. The UltraDeep survey strategy has only a ``Deep'' tier, converting the ``Medium'' tier from the Highz strategy into increased area for the Deep tier. While the UltraDeep strategy is not suggested as a replacement for existing \textit{Roman} SN survey strategies, it represents a useful class of survey that is worth investigating. 

\subsection{Generating Libraries of Mock Lensing Systems}
\label{sub:lens}
The work of \citet[][hereafter \citetalias{oguri_gravitationally_2010}]{oguri_gravitationally_2010} produced mock catalogs of strongly lensed quasars and SN for a series of current and future surveys. These included the \textit{Vera Rubin Observatory} \textit{Legacy Survey of Space and Time} (LSST) and the Joint Dark Energy Mission (JDEM), a concept telescope similar to the current \textit{Roman Space Telescope}. The \citetalias{oguri_gravitationally_2010} LSST catalog includes systems that produce lensed SN as doubly-imaged (doubles) and quadruply-imaged (quads), for a 10-year survey (in 3 month seasons) covering 20\,000 $\rm{deg}^2$. This procedure does not include image analysis, and therefore \citetalias{oguri_gravitationally_2010} assumed that lensed images are resolved if separated by more than $2/3~\theta_{psf}$, where $\theta_{psf}$ is the instrument PSF full width at half maximum (FWHM). 

\begin{table*}[ht]
\centering

\caption{\label{tab:rst_survey} Lensed SN yields for all of the surveys investigated in this work, described in Section \ref{sub:sn_surveys}.}
\begin{tabular}{ccrcrrrrrrr}
\toprule
\multicolumn{1}{c}{\textbf{Survey}}  &\multicolumn{1}{c}{\textbf{Tier}}  &\multicolumn{1}{c}{\textbf{Area}}&\multicolumn{1}{c}{\textbf{Filters}}&\multicolumn{1}{c}{\textbf{Cadence}} &\multicolumn{1}{c}{\textbf{Exp. Time}} & \multicolumn{1}{c}{\textbf{Visit Depth$^a$}}&\multicolumn{1}{c}{\textbf{H18 Yield}} &\multicolumn{3}{c}{\textbf{Lensed SN Yields}}  \\
 & &\multicolumn{1}{c}{$\rm{deg}^2$} & &\multicolumn{1}{c}{Days}& \multicolumn{1}{c}{s}  & \multicolumn{1}{c}{mag (band)}&\multicolumn{1}{c}{Ia} & \multicolumn{1}{c}{CC} &\multicolumn{1}{c}{Ia}& \multicolumn{1}{c}{Total}  \\
\hline
Lowz&&&&\\
&Shallow&142.3&RZYJ&5&13&22.0 (J)&6&<1&<1&<1\\
&Medium&66.91&RZYJ&5&67&24.8 (H)&4560&17&10&27\\
&\textbf{Total}&&&&&&4566&\textbf{17}&\textbf{10}&\textbf{28}\\
Allz&&&&\\
&Shallow&48.82&RZYJ&5&13&22.0 (J)&1225&<1&<1&<1\\
&Medium&19.75&RZYJ&5&67&24.8 (H)&5723&5&3&8\\
&Deep&8.87&YJHF&5&265&26.2 (H)&6640&12&6&18\\
&\textbf{Total}&&&&&&13588&\textbf{17}&\textbf{9}&\textbf{26}\\

Highz&&&&\\
&Medium&32.06&RZYJ&5&67&24.8 (H)&9354&8&5&13\\
&Deep&13.24&YJHF&5&265&26.2 (H)&9640&18&8&27\\
&\textbf{Total}&&&&&&18994&\textbf{27}&\textbf{13}&\textbf{40}\\
HLS&&&&\\
&--&2000&YJHF&$365^b$&300&26.4 (H)&--&\textbf{141}&\textbf{57}&\textbf{198}\\
UltraDeep$^c$&&&\\
&Deep&90.88&YJHF&10&265&26.2 (H)&--&\textbf{126}&\textbf{56}&\textbf{183}\\

\hline
\end{tabular}
\footnotesize
\begin{flushleft}

$^a$ These single visit depths are based on a $5\sigma$ detection threshold.

\

$^b$ The HLS survey is simulated using a single-epoch detection (requiring a follow-up strategy to obtain light curves). We estimate that the length of time between an initial visit and a ``discovery'' visit to the same field would be roughly 1 year.  The properties of the HLS are taken from the original SDT report \citep{spergel_wide_2015} and \citet{troxel_synthetic_2019}.

\

$^c$ This survey is not in \citetalias{hounsell_simulations_2018}; it is an alternative option explored here.

\end{flushleft}
\normalsize

\end{table*}

We follow the methodology of \citetalias{oguri_gravitationally_2010} to create similar catalogs of lensing systems based on the survey parameters presented in Section \ref{sub:sn_surveys} and Table \ref{tab:rst_survey}, with an update of the velocity function of galaxies (the distribution of galaxies as a function of velocity dispersion) that adds redshift evolution \citep{oguri_effect_2018}. As detailed in \citetalias{oguri_gravitationally_2010}, the velocity dispersion serves as a proxy for the total galaxy mass, used to define the lensing potential. We follow \citetalias{oguri_gravitationally_2010} by restricting our simulations in this work to galaxy-scale lenses, leaving predictions for group- and cluster-scale lenses to future work. This simplification enables simpler modeling of the lens parameters, and is reasonable as galaxy-scale lenses will dominate both the total lens population and the systems most useful for cosmology. Each survey in Table \ref{sub:sn_surveys} has a mock catalog produced that matches its specifications, so that the yields and properties of each strategy can be analyzed.

Each mock catalog of galaxies is generated following the assumed velocity function of galaxies, and we adopt a singular isothermal ellipsoid (SIE) with an external shear for the lensing potential of each galaxy. We use a Monte Carlo simulation to create each so-called ``macromodel'', containing properties of each lensing system (lens redshift, velocity dispersion, ellipticity, external shear, etc.) and each lensed source (source redshift, position, absolute magnitude, etc.) using the methodology described in \citetalias{oguri_gravitationally_2010}.
Next we solve the lens equation using the lens-modeling code {\tt glafic} \citep{oguri_mass_2010}, and if the source redshift and position result in multiple images, each image position, magnification, and time delay is added to the catalog. The final mock catalogs are 100 times over-sampled with respect to the estimated lensed SN yield, and only contain systems for which the peak brightness of the fainter (for doubles) or the third faintest (for quads) of the multiple images is above each catalog's detection threshold (see Section \ref{sub:survey_yields}). We note here that microlensing is neglected in the yield calculation for simplicity, as it is expected to have a small impact on detectability compared to time delay measurements and lens modeling \citep[e.g.,][]{dobler_microlensing_2006}. We check this approximation by calculating the change in the number of lensed SN ``detections'' due to the application of microlensing curves simulated in Section \ref{sec:sims}. We find the impact to be $\lesssim 4\%$ and that our yields here are relatively conservative, as the change leads to an increase in the lensed SN yield for Roman in line with the so-called ``magnification bias'' \citep[e.g.,][]{turner_statistics_1984,narayan_lectures_1997,barnacka_gravitational_2018}.

\citetalias{oguri_gravitationally_2010} adopted the condition that the peak magnitude distribution is cut off 0.7~mag brighter than
the deepest survey 10$\sigma$ single visit limiting magnitude. This was implemented so that the resulting lensed SN light curves are well sampled and suitable for time delay measurements. Here we are interested in the {\it total} detection rate for each survey strategy, recognizing that some discovered SN at the limit of the \textit{Roman Space Telescope} depth will not deliver useful light curves without additional follow-up observations. We therefore remove this restriction, and instead apply the 5$\sigma$ limiting magnitudes reported in Table \ref{tab:rst_survey} as peak magnitude cuts for each catalog. 

Using the parameters in Table \ref{tab:rst_survey}, we create a separate catalog of lenses for each SNIa survey strategy and each survey tier. The distributions of image separation vs. $\Delta t$ and $z_{lens}$ vs. $z_{src}$ for the Highz catalog are shown in Figure \ref{fig:survey_redshifts}, compared with the \citetalias{oguri_gravitationally_2010} LSST catalog. \textit{Roman} will have the ability to resolve much lower image separations with its PSF FWHM of $\sim0.11$ arcseconds. Longer time delays are also desirable to increase the relative precision of measurements (see Section \ref{sec:delays}), and the tail of the time delay distribution for \textit{Roman} extends further into the long time delay regime. \textit{Roman} will also be observing SN at significantly higher redshifts, which lengthens the observability period due to increased time dilation and extends the cosmological leverage of the overall lensed SN sample.

\subsection{Lensed SN Yields by Survey Strategy}
\label{sub:survey_yields}

We consider three of the \textit{Roman Space Telescope} SN survey strategies presented in \citetalias{hounsell_simulations_2018}, described in Section \ref{sub:sn_surveys} (Table \ref{tab:rst_survey}). The single visit depth for each tier is held constant across the \citetalias{hounsell_simulations_2018} survey strategies, while the survey area and number of filters are varied (see \citetalias{hounsell_simulations_2018} Table 7 for more details). In addition to these SNIa surveys from \citetalias{hounsell_simulations_2018} (which are investigated throughout the rest of this work), we calculate yields for the additional two surveys described in Section \ref{sub:sn_surveys}: the HLS and UltraDeep.

When predicting yields, we separate the discovered SN into SNIa, and core-collapse SN (CCSN). The CC class is further split into sub-classes in Section \ref{sec:sims} for light curve simulations. It should also be noted that the cadence for any of the surveys in Table \ref{tab:rst_survey} (apart from the HLS) can be adjusted for a linear impact on SN yield (i.e. doubling the cadence means $\sim$double the survey duration and therefore $\sim$double the yield). These changes in cadence have a minimal impact on precision ($<0.5$ days, Section \ref{sec:delays}), and can have a large impact on lensed SN yield. 

The mock lens catalogs created in Section \ref{sub:lens} are simulated to accurately reflect the expected number of observed lensed SN for each survey strategy, over-sampled by a factor of 100. We therefore report the yield expected for each survey strategy by simply dividing the number of lenses produced in each catalog by 100 (Table \ref{tab:rst_survey}). One could apply relative CCSN rates to the CCSN yield quoted here to determine lensed SN yield predictions for CCSN sub-types.

\section{Lensed Supernova Light Curve Simulations}
\label{sec:sims}
Here we describe the simulation procedure of a representative sample of SNIa and CCSN light curves used to determine the time delay precision obtainable with \textit{Roman Space Telescope}. For this precision analysis, we do not restrict ourselves to a lens catalog that follows one of the survey strategies from Section \ref{sub:sn_surveys}. Instead, we follow the methodology in Section \ref{sub:lens} to create a vastly over-sampled catalog that fully covers the diversity of lens parameters expected for each SN type. This ``light curve simulation lens catalog'' is simulated to a depth $0.6$ magnitudes beyond the deepest limiting magnitude of \textit{Roman} over an area $\sim15$ times larger than the widest survey strategy. Next, we apply the methodology of \citetalias{hounsell_simulations_2018} to simulate the lensed SN light curves and determine which would be detected by the \textit{Roman Space Telescope}.  Therefore the final sample is based on realistic detection efficiencies and selection requirements. The light curve simulation lens catalog is used throughout the remainder of this section, using a random sample of 100 lenses across all SN types. The following is an outline of the process described in this section for reference:
\begin{enumerate}
    \item \textbf{Create light curve simulation lens catalog}--This lens catalog is a highly over-sampled and deeper version of the lens catalogs in Section \ref{sub:lens} specifically for the time delay precision analysis
    
    \item \textbf{Simulate SN light curves}--Use \citetalias{hounsell_simulations_2018} ``Imaging:Allz'' survey strategy to define cadence, depth, etc., then combine with \textit{Roman Space Telescope} properties defined in \citetalias{hounsell_simulations_2018} using {\fontfamily{qcr}\selectfont{SNANA}} package (see Section \ref{sub:snana}) to create a realistic SN light curve for each lensed image
    \item \textbf{Match simulated SN to simulated lenses}--Attempt to match each ``observed'' \textit{Roman} SN with a lens from our catalog that has the same source redshift, oversampling the \textit{Roman} survey until we have a sample of 50 lenses for each SN type with 10 distinct light curves per lens (Table \ref{tab:glsn_sim}, Section \ref{sub:snana})
    \item \textbf{Microlensing simulations}--Create 12 distinct sets of  microlensing maps based on different choices of mass model, and match each image from every lens with a unique map (Section \ref{sub:micro})
    \item \textbf{Add microlensing to light curves}--Convolve 100 randomly located SN profiles with every microlensing map for each of the 12 mass models, creating 1200 microlensing variations for each simulated SN (Table \ref{tab:glsn_sim}). Apply each microlensing curve to the baseline light curve (Section \ref{sub:final-sim})
    \item \textbf{Selection Cuts}--Apply selection cuts on the microlensed light curves, removing those that would not have been detected (5$\sigma$ detection)
\end{enumerate}

\begin{table*}[ht!]
\centering
\caption{\label{tab:glsn_sim} The number of lensed SN light curves simulated for this work.}
\begin{tabular*}{\textwidth}{@{\extracolsep{\stretch{1}}}*5{r}}
\toprule
 \multicolumn{1}{c}{\textbf{Sample}}&\multicolumn{1}{c}{\textbf{N Lenses}} &\multicolumn{1}{c}{\textbf{N LC/Lens}} &\multicolumn{1}{c}{\textbf{N $\mu-$Lens}}&\multicolumn{1}{c}{\textbf{Total Simulations}} \\

\hline
Each SN Type&50&10&1200&600,000\\
Combined&200&-&-&2,400,000\\
\end{tabular*}

\end{table*}

\subsection{Macrolensed Light Curves with SNANA}
\label{sub:snana}

We adopted the methodology of \citetalias{hounsell_simulations_2018} to simulate the \textit{Roman Space Telescope} SN surveys, taking into account the most significant uncertainties to create realistic light curves and survey properties using the SuperNova ANAlysis ({\fontfamily{qcr}\selectfont{SNANA}}) simulation package \citep{kessler_snana:_2009,kessler_first_2019}. {\fontfamily{qcr}\selectfont{SNANA}} simulates SN light curves for an arbitrary set of survey properties while accounting for variations in noise, atmospheric transmission, cadence, and other telescope properties. Due to its speed, accuracy, and flexibility, {\fontfamily{qcr}\selectfont{SNANA}} has become the standard tool for simulating SN surveys in recent years \citep[e.g.,][]{betoule_improved_2014,scolnic_complete_2018,smith_first_2020}. We provide {\fontfamily{qcr}\selectfont{SNANA}} with the same information about \textit{Roman} as did \citetalias{hounsell_simulations_2018} (e.g. filter properties, noise sources, etc.), in order to create similarly realistic SN light curves and host properties. 
We simulate the population of SNIa using the extended SALT2 model \citep{guy_salt2:_2007,pierel_extending_2018}, and populations of core-collapse SN using a library of SED time series (see the Appendix, Table \ref{Atab:snana_seds}).

For the work that follows, we divide the core collapse SN (CCSN) population into three sub-classes: Type Ib/c, Type II, and Type IIn.   
The Type Ib/c sub-class includes all Hydrogen-poor ``stripped-envelope'' CCSN \citep{modjaz_new_2019}.
We have similarly combined both the ``plateau'' Type II-P and ``linear'' II-L into a single ``Type II'' subclass, following recent work on the subject \citep[e.g.,][]{anderson_characterizing_2014,sanders_toward_2015,valenti_diversity_2016}.  Although we are not considering spectroscopy in this work, we segregate the Type IIn---distinguished by the presence of narrow spectral line components---because this sub-class includes many relatively bright CCSN \citep{kiewe_caltech_2012} that will be over-represented in future lensed SN samples \citep{wojtak_magnified_2019}.

Additional functionality has been added to the {\fontfamily{qcr}\selectfont{SNANA}} simulation for this (and future) work, which enables the creation of multiple magnified light curves of the same SN based on a library of lenses provided by the user (see Section \ref{sub:lens}). When a SN is generated, it is matched to a randomly generated lensing system from the lens library with the same source redshift. For each lensing system image, the simulation creates an identical light curve (with a unique noise realization) and applies the associated macrolensing magnification and time delay. This capability provides realistic light curves for resolved, strongly lensed SN. With this new software in place, our {\fontfamily{qcr}\selectfont{SNANA}} simulations proceed in the following manner:
\begin{enumerate}
    \item \textbf{Source Model}
    \begin{enumerate}
        \item Generate source SED at each epoch determined by survey strategy and SN type
        \item Apply cosmological dimming, weak lensing, peculiar velocity, and redshift SED
        \item Apply galactic extinction 
        \item Integrate redshifted SED for each filter to create (true) photometric light curve
    \end{enumerate}
    \item \textbf{Strong Lensing}
    \begin{enumerate}
        \item Attempt to match SN with lens in catalog based on source redshift
        \item If matched, create N identical light curves of SN based on lens parameters
        \item Apply time delay and macro-magnification to each light curve based on matched lens
    \end{enumerate}
    \item \textbf{Noise Model}
    \begin{enumerate}
        \item Convert true light curve magnitudes to true flux and uncertainty using estimated PSF, sky-noise, and zero-points for \textit{Roman}
        \item Apply Poisson noise to get measured flux in photoelectrons
    \end{enumerate}
    \item \textbf{Trigger Model}
    \begin{enumerate}
        \item Check for detection based on 1$\sigma$ SNR in 2+ bands (more stringent cuts follow in Section \ref{sub:final-sim})
        \item Write selected events to data files, linking multiply-imaged light curves with a unique lens ID
        
    \end{enumerate}
\end{enumerate}

Employing this simulation process, we use the ``Imaging:Allz'' survey strategy from Table \ref{tab:rst_survey} and simulate light curves from the Shallow, Medium, and Deep Tiers. We remove light curves that do not meet a minimum threshold of peak signal-to-noise ratio (SNR) of 1 in at least 2 filters. This initial threshold is intentionally very low, because microlensing effects may amplify the signal of an otherwise undetectable SN. Once microlensing is added in Section \ref{sub:micro}, we perform a more stringent cut by only accepting light curves with peak SNR$\geq5$ in at least 1 filter. Next, we significantly over-sample the SN survey such that we obtain 10 unique realizations for 50 discoverable lensed SN, which ensures a diverse sampling of light curve templates (CC) or light curve parameters (Ia). Finally, each lensed SN light curve is varied a further 1,200 times by different microlensing models and microcaustic convolutions (see Section \ref{sub:micro}). Each lens therefore has 12,000 unique SN light curve realizations, for a total of 600,000 light curve simulations per SN type. These light curves cover a complete range of the microlensing parameter space corresponding to that lens, and a variety of possible light curves for each SN type. Table \ref{tab:glsn_sim} shows the breakdown of ``discovered'' lenses for each simulated SN type, and the resulting number of light curves produced in this section (2.4 million).  The number of lenses per SN type is chosen to balance a sufficient sampling of lensing parameter space with computational efficiency, while the number of light curves per lens was chosen to ensure no single light curve realization biases our results. For an explanation of the number of microlensing variations per SN, see Section \ref{sub:micro}. The combined sample contains over 2 million simulated light curves, which will be made public for future work.

\begin{figure}[h!]
\centering
\includegraphics[trim={1.2cm 1.2cm 2cm 3cm},clip, width=0.45\textwidth]{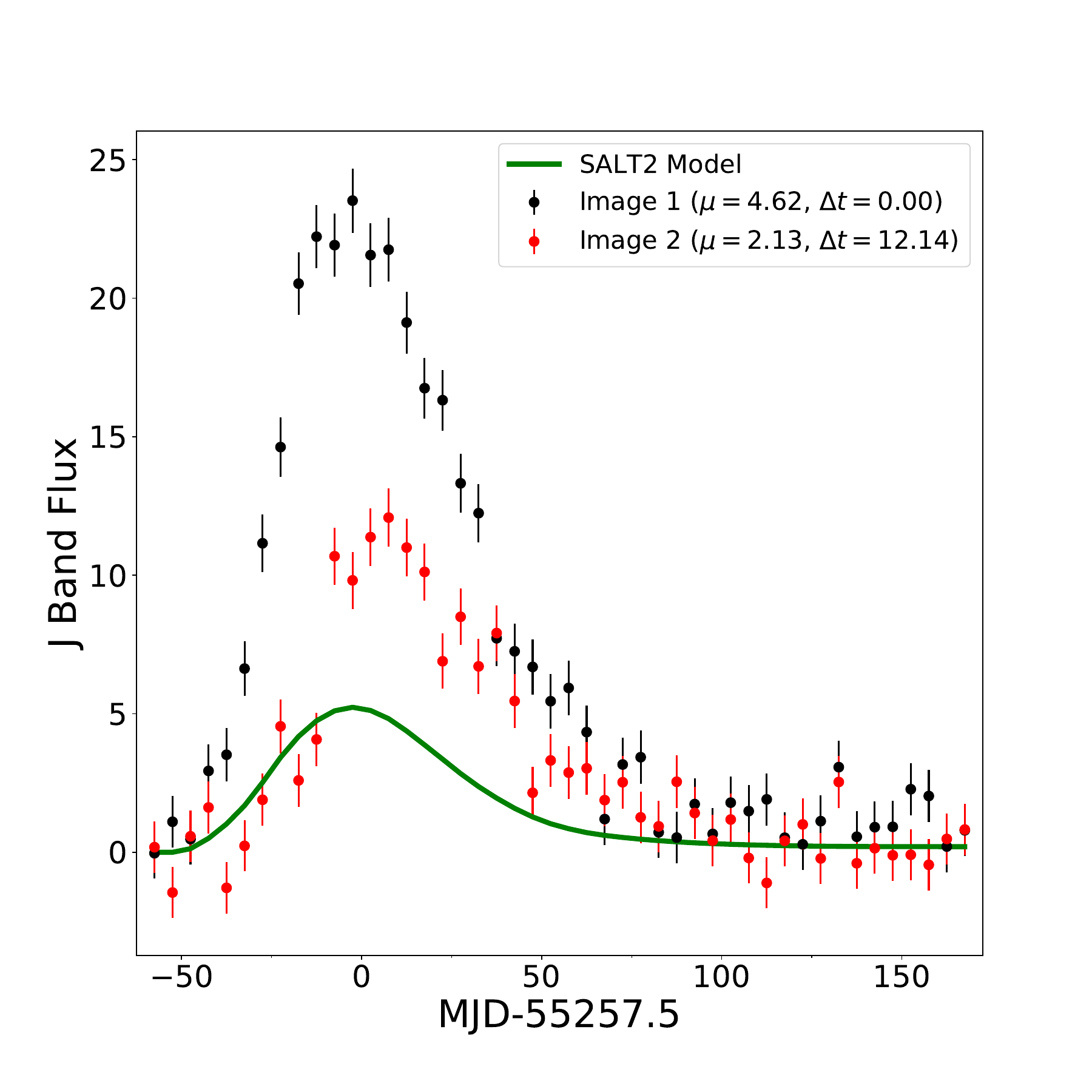}
\caption{\label{fig:snana_lc} An example of a lensed SNIa simulated from {\fontfamily{qcr}\selectfont{SNANA}}, before microlensing is applied in Section \ref{sub:micro}. The x-axis shows observer-frame time relative to peak brightness. The solid green line shows the underlying (un-lensed) SALT2 model that was used to generate the observed light curve. The two simulated SN image light curves, with unique (background-limited) noise realizations based on the telescope and survey properties applied, are shown as points with error bars. Image 1 (black) has a magnification of 4.62, and image 2 (red) has a magnification of 2.13, with a time delay of 12.14 days.}
\end{figure}

\subsection{Microlensing Simulations}
\label{sub:micro}

When light rays from the expanding photosphere of a SN pass within the lensing potential of a set of stellar mass objects in the lens plane, fluctuations in the light curve can be introduced on timescales of weeks to months \citep{dobler_microlensing_2006}.
This effect, given the name of microlensing due to the size of the Einstein radii associated with these deflectors (see equation \ref{eq:einstein} below), is a critical component of realistic lensed SN simulations \citep{dobler_microlensing_2006,foxley-marrable_impact_2018,goldstein_precise_2018,bonvin_impact_2019,pierel_turning_2019,huber_strongly_2019}.

The main parameters defining the microlensing properties are the values of the local convergence, $\kappa$, and shear, $\gamma$, fields, which are derived from the lens macromodel, at the location of the lensed SN macro-images (see Section \ref{sub:lens}).
The convergence causes a focusing of light rays, leading to an isotropic distortion of a source, while the shear introduces an anisotropic distortion. This shape modification usually increases (but may decrease) the surface area of the source while conserving surface brightness producing a magnification of its total observed flux \citep{narayan_lectures_1997}. The $\kappa$ parameter combines the contribution to the convergence from two mass components in the primary deflector: compact matter, $\kappa_*$, and smoothly distributed matter, $\kappa_s$, where $\kappa=\kappa_*+\kappa_s$. The parameter $s\equiv\kappa_s/\kappa$ defines the relative fraction of convergence from smoothly distributed mass components, such as dark matter, compared to the total, which includes grainy components, such as point-mass microlenses.

In our simulations, $\kappa$ is due only to the primary deflector, modeled as an SIE, while the shear includes an external contribution due to the net effect of all other masses along the line of sight.
To model the compact matter component of $\kappa$, and hence derive $s$, we use the
\begin{figure*}[ht!]
\centering
\includegraphics[trim={1cm .5cm .5cm .5cm},clip, width=\textwidth]{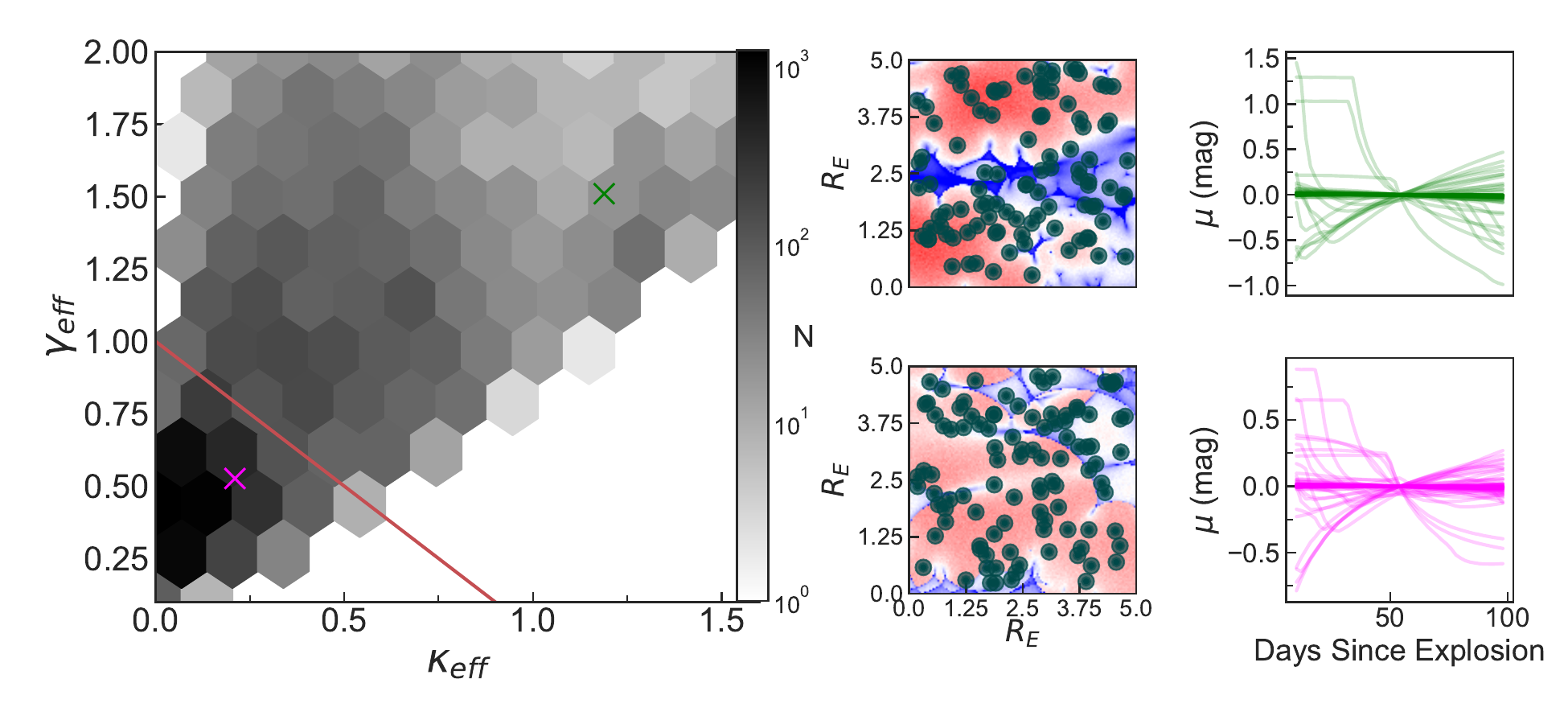}
\caption{\label{fig:micro_overview} Examples of microlensing curve creation for two regions of the parameter space. Left: The effective convergence ($\kappa_{eff}$) and shear ($\gamma_{eff}$, see Section \ref{sub:micro_precision}) for all unique microlensing realizations produced in this work, separated by the critical line (red solid line, separating the saddle point and minimum regions above and below, respectively). The magenta X is a more probable microlensing realization, the green X is a less probable realization. Center: The microcaustics corresponding to each scenario, with 100 random SN placements represented by green discs. Right: The resulting microlensing curves for each scenario, making it apparent that higher values of $\kappa_{eff},\gamma_{eff}$ lead to more extreme microlensing regimes (note the difference in magnification scale between the upper and lower panels). }
\end{figure*}
approach provided in \citet{vernardos_microlensing_2019}\footnote{The reader is referred to that work for the detailed derivation and equations.}: an elliptical S\'{e}rsic profile \citep{sersic_influence_1963} that has the same orientation and ellipticity as the SIE is used to describe the stellar mass in the lens galaxy.
The profile is scaled by the effective radius, $\theta_{\rm eff}$ (in arcsec), and normalized using the central dark matter fraction.
For the former, we use a fitted relation to the measured $\theta_{\rm eff}$ and $\theta_{\rm Ein}$ (see equation \ref{eq:einstein}) of the SIE from the SLACS \citep{auger_sloan_2009} and CASTLES \citep{oguri_stellar_2014} lens samples, while for the latter we use fits from \citet{auger_sloan_2010} based on the Chabrier or Salpeter Initial Mass Function (IMF).
Together with three values that we examine for the S\'{e}rsic index, $n=2,4,7$, we end up with 12 different ways to couple the compact to the total mass distribution in the lens galaxy and obtain a value for $s$ at the location of the multiple SN images.

Based on the values of $\kappa,\gamma,s$, which define the overall properties of each microlensing configuration, we create a realization of microlenses randomly distributed on the lens plane and solve the lens equation \citep[e.g. see Equation 17 and Fig. 2 in][]{schmidt_quasar_2010}.
However, instead of finding all the $2N+1$ micro-images produced by $N$ microlenses \citep{burke_multiple_1981}, which are anyway unobserved, we compute the value of the magnification on a pixellated grid on the source plane, viz., a magnification map.
We use the high-quality magnification maps from the comprehensive GERLUMPH database\footnote{\href{http://gerlumph.swin.edu.au/}{http://gerlumph.swin.edu.au/}}, as it already covers the parameter space we require for our light curve simulations \citep{vernardos_gerlumph_2014,vernardos_gerlumph_2015}. 
\begin{figure*}[ht!]
\centering
\includegraphics[trim={1.1cm 1.1cm 0cm 0cm} ,clip,width=\textwidth]{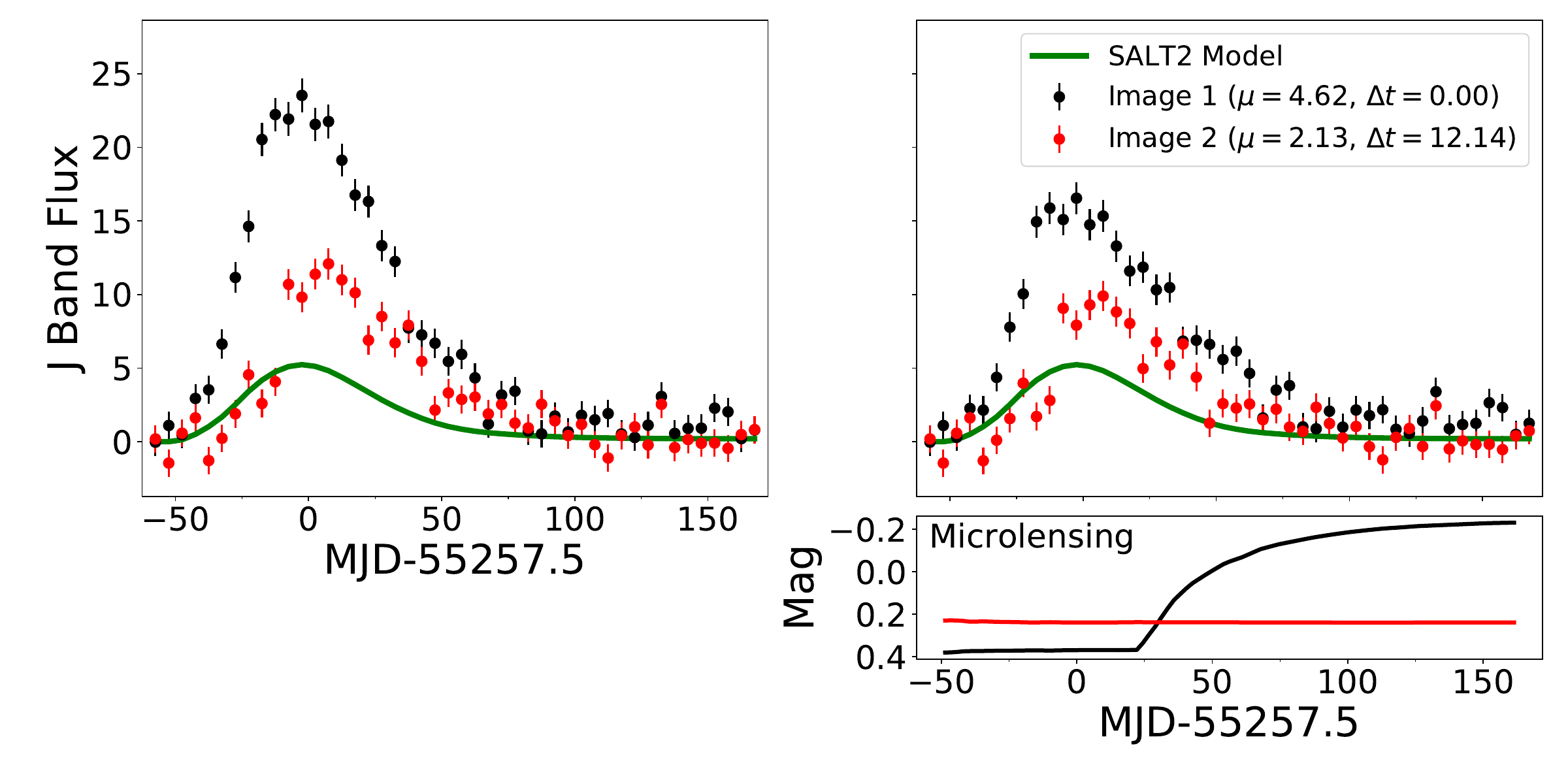}
\caption{\label{fig:microlensed} Left panel: Simulated lensed SN from Figure \ref{fig:snana_lc}, for reference. Upper right panel: Simulated lensed SN from Figure \ref{fig:snana_lc}, now with microlensing applied from Section \ref{sub:micro}. Lower right panel: The microlensing curve applied to image 1 (black), and image 2 (red), in magnitudes. In both light curve plots, the solid green curve denotes the underlying (un-lensed) model used to create the simulation before the application of microlensing.}

\end{figure*}
The GERLUMPH maps also have a sufficiently high resolution (10,000 pixels) and width (25 Einstein radii, $R_E$) for our purposes.
The characteristic length scale of microlensing is the Einstein radius, $R_E$, whose projected value on the source plane for a point-mass lens is equal to:
\begin{equation}
    \label{eq:einstein}
    R_E=\sqrt{\frac{4GM}{c^2}\frac{D_{ls}D_s}{D_l}},
\end{equation}
where $D_s, \ D_l, \ \rm{and} \ D_{ls}$ define the angular diameter distances to the source, lens, and between the lens and source, respectively, $M$ is the mean mass of the microlenses, $G$ the gravitational constant, and $c$ the speed of light.
Finally, the GERLUMPH maps adopt the well-established simplification for the microlens mass function, whereby all microlens masses are set to $1M_\odot$ \citep[e.g.,][]{wambsganss_probability_1992,schechter_qualitative_2004}. If the source has a finite size, larger than the magnification map pixels, as is the case for an expanding SN, a convolution between the map and the SN brightness profile is performed to extract the value of the microlensing induced magnification.
Following the methodology of \citet[][hereafter \citetalias{pierel_turning_2019}]{pierel_turning_2019} and \citet{foxley-marrable_impact_2018}, we create a model for an expanding SN photosphere as an achromatic flat-disk brightness distribution.
By convolving the expanding photospheric model with the magnification map, we obtain a microlensing magnification curve (Figure \ref{fig:micro_overview}). The SN is placed randomly at 100 locations per magnification map to probe regions that may have varying magnification properties and to create a statistically representative sample. We found that 100 samples is small enough to not over-sample the region with overlapping profiles, but large enough to explore the range of curves possible for a given microlensing map. These 100 random locations for each of the 12 compact mass models described above, form the 1,200 microlensing variations referenced in Section \ref{sub:snana} and Table \ref{tab:glsn_sim}.

We have made three simplifying approximations here: (1) the disk has no transverse velocity across the map, (2) projecting the 3-D 
photosphere into a flat disk, and (3) assuming no intensity variation across the disk (i.e. achromaticity). Previous work has found that, depending on the lens and source redshift ratio, the transverse velocity of the source profile across the magnification map should be a few hundred to a few thousand km~s$^{-1}$ in extreme cases. The velocity of the expanding photosphere of each SN profile is $>10^4$~km~s$^{-1}$, so this approximation is reasonable \citep{zheng_empirical_2018,pierel_turning_2019}.
The flat disk approximation has been shown to be sufficiently accurate up to an ignored time delay measurement bias on the order of 0.1 days \citep[the ``microlensing time delay''][]{bonvin_impact_2019}.
However, a potentially significant improvement would be to adopt a wavelength-dependent 2-D projected specific intensity profile, as used by \citet{goldstein_precise_2018} and \citet{huber_strongly_2019}. 
Such an intensity profile would better represent a SN atmosphere, which emits light varying in wavelength and brightness as a function of radius from the center  \citep[e.g.,][]{kasen_timedependent_2006} in such a way that chromatic microlensing effects are often predicted \citep{goldstein_precise_2018,huber_strongly_2019}.
We defer this improvement to future work, as such profiles are not available for all SN types of interest and the most  

\

\

\

\noindent important contribution to microlensing effects remains the size of the lensed object \citep[i.e. quasar or SN photosphere, see][]{mortonson_size_2005,vernardos_microlensing_2019}.

\subsection{Final Simulated Lensed Supernova Light Curves}
\label{sub:final-sim}
We now have a large repository of simulated (macro-) lensed light curves for each SN type in Table \ref{tab:glsn_sim}, as well as 1200 unique achromatic microlensing curves for each SN. We found in the course of our simulations that microlensing can vary the flux of a given light curve by over a magnitude in some cases (Figure \ref{fig:micro_overview}). While these cases are rare, it does introduce the possibility that a lensed SN in our sample could be pushed above or below the detection threshold, owing to the microlensing alone. We account for this possibility by calculating the time-varying change in flux due to microlensing and updating both the ``observed'' flux and the flux uncertainty in each data point accordingly. The random scatter that drives an observed flux value away from the generative model is dominated by the sky noise, but also includes a Poisson noise term.  Whenever microlensing leads to an increase in flux, the Poisson noise necessarily increases, and we therefore apply the updated Poisson noise, resulting in additional scatter in the ``observed'' flux. However, in cases where microlensing causes de-magnification (thus decreasing the Poisson noise), we do not change the value assigned for the observed flux. This simplification is used because of the difficulty inherent in removing a randomly drawn scatter term after it has been applied. As the changes in flux from the additional Poisson noise due to microlensing are in general $<1\%$, this simplification will not lead to any significant bias in the final selection cuts.

After microlensing has been applied to each light curve and the flux uncertainties have been adjusted, we apply a selection cut that determines the detectability of each lensed SN. This final cut removes all simulated SN that do not have at least 1 image with a light curve peak SNR of $\geq5$ in 1 or more filters, and we ensure that our final sample matches the description found in Table \ref{tab:glsn_sim}. As an example, the macro-lensed light curve from Figure \ref{fig:snana_lc} is shown again here in Figure \ref{fig:microlensed}, now with a random microlensing curve applied. A more selective cut could be used at this stage, which would reduce the lensed SN sample in favor of higher precision time delay measurements. Our choice of a cutoff at SNR$\geq5$ in a single filter matches our yield calculations performed in Section \ref{sec:NGRST}.

\section{Measuring Time Delays}
\label{sec:delays}
In this section we fit each of the simulated light curves from section \ref{sec:sims} to estimate the time delay precision measurable for each SN type. Additionally, we discuss the topic of microlensing and its impact on time delay measurements and SNIa standardizability for the \textit{Roman Space Telescope}.

\subsection{The {\fontfamily{qcr}\selectfont{SNTD}} Software Package}
\label{sub:sntd}
We use version 2 of the SuperNova Time Delays ({\fontfamily{qcr}\selectfont{SNTD}}\footnote{\href{https://sntd.readthedocs.io/en/latest/index.html}{https://sntd.readthedocs.io}}) package from \citetalias{pierel_turning_2019} to measure the time delays of all lensed SN simulated in Section \ref{sec:sims}. {\fontfamily{qcr}\selectfont{SNTD}} is designed to take advantage of the fact that the multiple light curves of a lensed SN are created from the same SN explosion, which means they can be represented by a single light curve model. Since we have light curve templates for a variety of SN types (see Section \ref{sub:snana}), {\fontfamily{qcr}\selectfont{SNTD}} is capable of fitting multiple light curves that are incomplete due to the observing strategy. By combining the multiple light curve fits, {\fontfamily{qcr}\selectfont{SNTD}} can determine the intrinsic (e.g. color, shape) and lensing (e.g. magnification, time delay) SN parameters. 

The {\fontfamily{qcr}\selectfont{SNTD}} software package has three separate methods for measuring time delays using light curves from resolved images of lensed SN: {\it Parallel}, {\it Series}, and {\it Color} (for details on these methods, see \citetalias{pierel_turning_2019}). In cases where the light curves are relatively well-sampled and have been observed before and after peak brightness (as is the case here), the \textit{Parallel} method is just as effective as the \textit{Series} method and is more computationally efficient. We note that the Color method might be more effective for measuring time delays in the case of chromatic microlensing as it determines the time delay by way of the SN color curve instead of its multi-band light curve \citep{goldstein_precise_2018}. In our work all simulated microlensing is achromatic (see Section \ref{sub:micro}), so the {\it Parallel} method is used to establish the time delay precision for \textit{Roman}, which fits each image's light curve separately but combines fitting posteriors to jointly estimate light curve parameters. Note that this method does not attempt to algorithmically remove microlensing effects from the light curves, as attempting to do so can cause an unintended bias in measured time delays \citep[e.g.,][]{tewes_cosmograil_2013,pierel_turning_2019}. Instead, it relies on the relative inflexibility of the template-based fitting approach compared to flexible functions \citep[e.g.,][]{rodney_sn_2016,huber_strongly_2019} to limit the size of the global bias in time delay results \citep[see Section \ref{sub:delays};][]{pierel_turning_2019}. We still evaluate the impact of this choice in Section \ref{sub:fit_micro}.

\pagebreak
\subsection{Time Delay Measurements}
\label{sub:delays}
We fit each light curve from Section \ref{sec:sims} using the methodology outlined in Section \ref{sub:sntd}, which requires a SN model or template for each fit. Type Ia SN are fit with the SALT2 model \citep{guy_salt2:_2007}, extended into the NIR \citep{pierel_extending_2018}. Type IIP and IIn (Ibc) are fit with all Type II (Ibc) SN templates in the SNCosmo Python package\footnote{\href{https://sncosmo.readthedocs.io/en/v2.1.x/}{https://sncosmo.readthedocs.io}} (Table \ref{Atab:snana_seds}), except for the template SED time series used to simulate that SN (if present) to avoid biasing the precision results. The ``best-fit'' template for each CCSN is determined by comparing the Bayesian Evidence across all template fits. 

\begin{figure}[b!]
\centering
\includegraphics[trim={2cm 1.5cm 1cm 3.5cm},clip, width=0.5\textwidth]{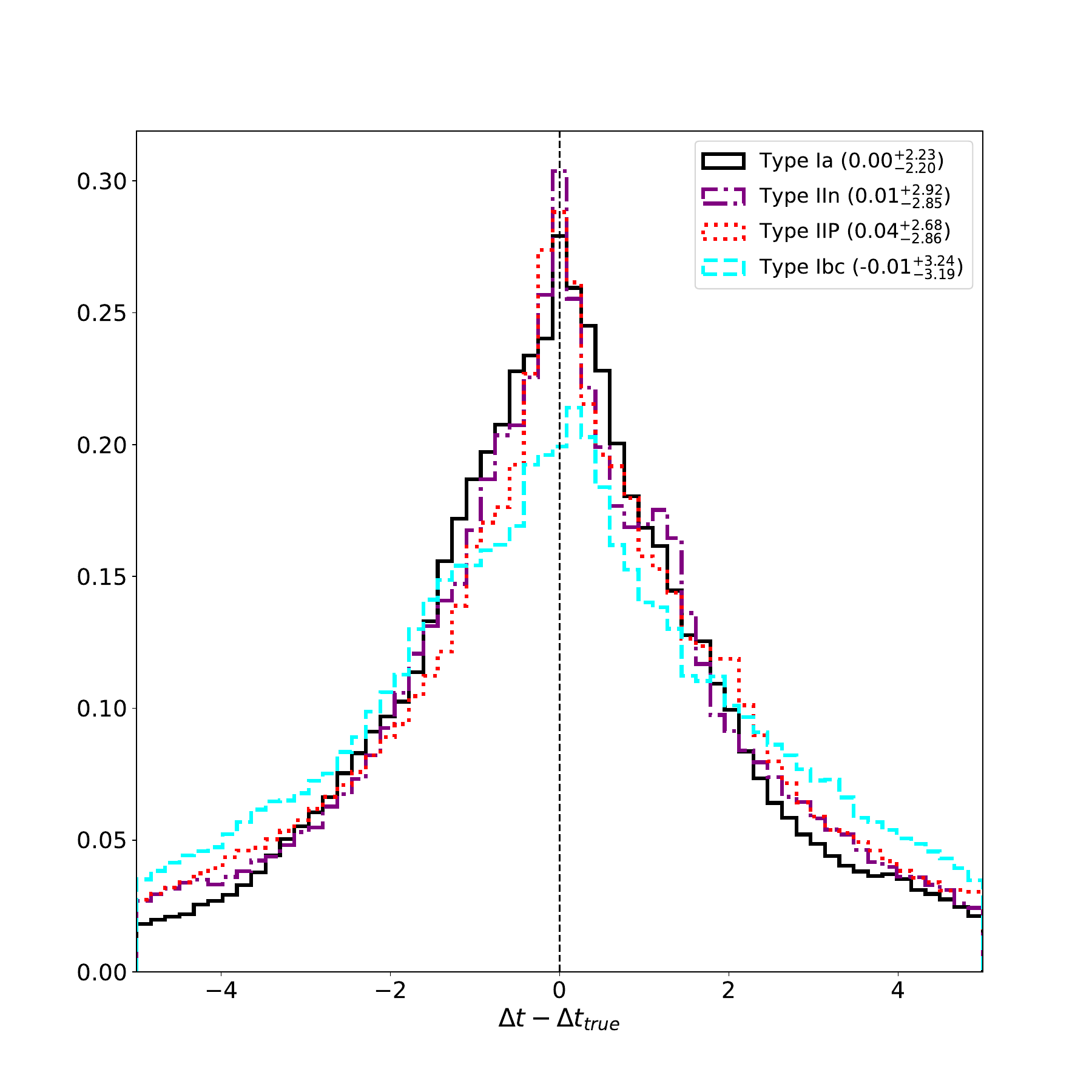}
\caption{\label{fig:precision_by_type}Distributions of measured time delays for each SN type. All SN types have a slight asymmetry apparent in the legend. There are small biases in accuracy across all CCSN types (which are statistically significant), but SNIa measurements are unbiased. }
\end{figure}

We find statistically significant biases in time delay accuracy for all CCSN types, but not for SNIa (Table \ref{tab:prec_types}). However, for CCSN we measure a bias of $\sim0.03$ days, which amounts to $<1\%$ for $\sim80\%$ of the CCSN in our catalogs. The precision of time delay measurements varies from $2-3$ days depending on SN type. Due to the non-Gaussian nature of the residual time delay distributions, the width of each distribution (our reported statistical precision) is estimated by the 16th and 84th percentile ranges. Unsurprisingly due to their brightness and homogeneity, SNIa produce the most precise (and accurate) time delay measurements, which are $\sim25-50\%$ more precise than CCSN time delay measurements (Table \ref{tab:prec_types}, Figure \ref{fig:precision_by_type}). 
\begin{table}[t!]
\centering
\caption{\label{tab:prec_types} The bias and precision for time delay measurements performed for each SN type in this section.}
\begin{tabular*}{\columnwidth}{@{\extracolsep{\stretch{1}}}*{3}{r}}
\toprule
\multicolumn{1}{c}{\textbf{SN Type}} & \multicolumn{1}{c}{\textbf{Bias}}& %
\multicolumn{1}{c}{\textbf{Precision}} \\
 &\multicolumn{1}{c}{Days}&\multicolumn{1}{c}{Days} \\
\hline
Ia&$0.00\pm+0.003$&$2.22$\\
Ibc&$-0.01\pm0.004$&3.22\\
IIP&$0.04\pm0.004$&2.77   \\
IIn&$0.01\pm0.004$&2.89\\
\end{tabular*}

\end{table}
Although the light curves simulated in Section \ref{sec:sims} used the ``Allz'' strategy, the quality of light curves is not expected to vary significantly for different survey strategies as the exposure time and cadence are fixed across all the survey variants. The exception to this expectation is UltraDeep, which has half the number of observations compared with existing SNIa survey strategies. However, we find that doubling the cadence from 5 days to 10 days only reduces the measured time delay precision by $\lesssim0.5$ days, and in general the deep imaging results in more precise light curve photometry. Since UltraDeep is not a proposed SNIa survey strategy, we assume here that these effects roughly cancel and no precision is lost or gained for the UltraDeep strategy. Therefore, we consider these precision results throughout the rest of this paper as constant across any \textit{Roman} survey, and use this information to compare the value of each strategy for SN time delay cosmography. 
\begin{figure}[t!]
\centering
\includegraphics[trim={1.25cm .5cm .5cm .5cm},clip,width=0.48\textwidth]{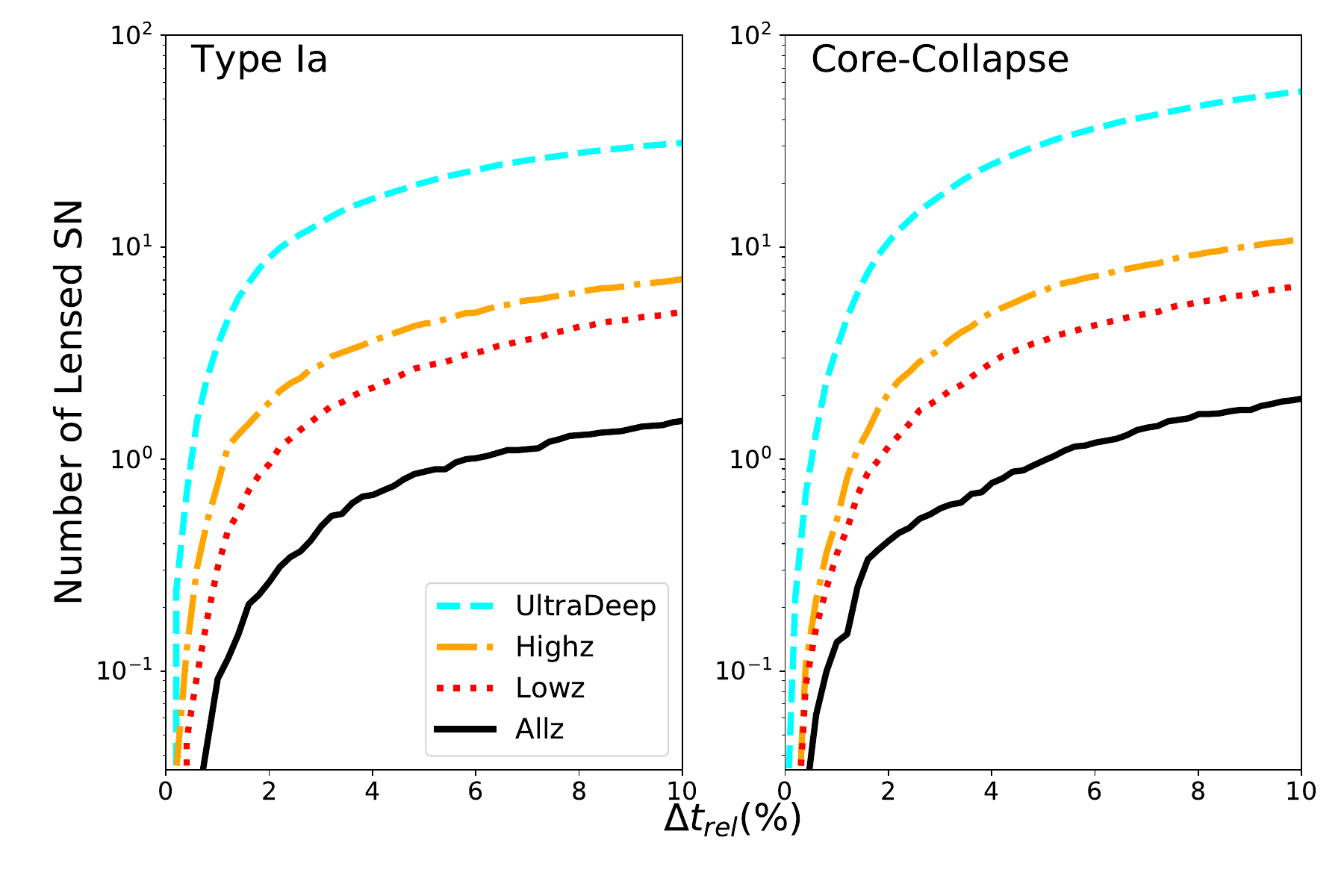}
\caption{\label{fig:survey_precision} A visual representation of Table \ref{tab:survey_prec_nums}. The left and right panels describe the SNIa and CCSN populations of each survey strategy, respectively. The x-dimension of each plot is the relative time delay precision $\Delta t_{rel}$. The y-dimension is the cumulative sum of SN for a given precision, identifying how many SN of a given type at a specific time delay precision could be discovered by \textit{Roman}, for each SNIa survey strategy.}
\end{figure}

We use the rates measured in Section \ref{sub:survey_yields} to define a random sampling of lenses from each survey catalog, and divide the {\it absolute} $\Delta t$ precision (in days) shown in Table \ref{tab:prec_types} by the time delay of each lens to define the {\it relative} time delay precision as a percentage ($\Delta t_{rel}$).  It is this relative precision that propagates through to the uncertainty of the cosmological parameters, such as $H_0$.  Here $\Delta t_{rel}$ for CCSN is the mean of the precision found for each CCSN type, weighted by their relative rates \citep{li_nearby_2011,rodney_type_2014}. Table \ref{tab:survey_prec_nums} reports our 
estimate for the number of SNIa and CCSN that could be discovered for each survey strategy as a function of $\Delta t_{rel}$.  Numbers in Table \ref{tab:survey_prec_nums} are cumulative, meaning that, e.g., the count with precision better than 5\% includes those with precision better than 2\% and 1\%.
Figure \ref{fig:survey_precision} presents this information graphically, showing the cumulative distribution curves as a function of $\Delta t_{rel}$.
\begin{table*}[t!]
\caption{ \label{tab:survey_prec_nums} Projected number of lensed SN for which $\Delta t$ can be measured with a given precision.}
\centering
\begin{tabular*}{\textwidth}{@{\extracolsep{\stretch{1}}}*{11}{r}}
\toprule
 & \multicolumn{10}{c}{\textbf{Number of SN with $\Delta t_{rel}$ better than...}} \\
  & \multicolumn{5}{c}{\textbf{(Core-Collapse)}} &\multicolumn{5}{c}{\textbf{(Ia)}} \\
\multicolumn{1}{c}{\textbf{Survey}} & \multicolumn{1}{c}{\textbf{1\% }} &\multicolumn{1}{c}{\textbf{2\%}} &\multicolumn{1}{c}{\textbf{5\%}} &\multicolumn{1}{c}{\textbf{10\% }} &\multicolumn{1}{c}{\textbf{CC Total}}&\multicolumn{1}{c}{\textbf{1\% }} &\multicolumn{1}{c}{\textbf{2\%}} &\multicolumn{1}{c}{\textbf{5\%}} &\multicolumn{1}{c}{\textbf{10\%}} &\multicolumn{1}{c}{\textbf{Ia Total}} \\
\hline
Lowz&<1&1&3&6&17&<1&1&2&4&10\\
Allz&<1&<1&1&2&17&<1&<1&1&1&9\\
Highz&1&2&6&9&27&1&2&4&6&13 \\
UltraDeep&5&11&27&47&126&5&9&19&29&56\\
\end{tabular*}
\end{table*}

These results vary across survey options, but we find that the Highz strategy from \citetalias{hounsell_simulations_2018} would be able to measure $\gtrsim30\%$ of its SNIa and $\sim25\%$ of its total SN with $5\%$ precision or better, with samples of $\sim4$ and $\sim10$ respectively. While the UltraDeep survey is not a proposed alternative to the strategies presented in \citetalias{hounsell_simulations_2018}, it shows the value that such a deep imaging survey would have for SN time delay cosmography. A survey like UltraDeep would be expected to yield more than 5 SN with $1\%$ precision, and 46 with $5\%$ precision (Table \ref{tab:survey_prec_nums}). 

\subsection{Characterizing Microlensing}
\label{sub:fit_micro}
We now explore the impact of microlensing on the light curve fitting performed in Section \ref{sub:delays}. We also evaluate our ability to identify and account for microlensing, as our fitting method does not attempt to directly remove microlensing from the light curve itself. We conclude with an estimate of the impact microlensing has on the standardizability of SNIa, which has been proposed as a method to help mitigate the effects of the mass-sheet degeneracy \citep{kolatt_gravitational_1998,holz_seeing_2001,oguri_gravitational_2003,rodney_illuminating_2015,xu_lens_2016}. 

\

\

\subsubsection{Effect on Accuracy and Precision}
\label{sub:micro_precision}
We perform two separate tests in this section. First, we ask how much the introduction of microlensing affects the accuracy and precision of $\Delta t$ measurements. Using SNIa as a case study, we fit all simulated light curves from Section \ref{sec:sims} before the application of microlensing and find a precision of 1.6 days with no significant bias. Comparing these fitting results to those presented in Section \ref{sub:delays} (with microlensing), we conclude that the global effect of microlensing for the SNIa in this work is to degrade the time delay precision by $\sim$0.6 days, while (achromatic) microlensing leaves SNIa time delay accuracy unbiased. However, we note that a bias on time of peak measurements for individual images does emerge, and is $\sim$0.1 days. These results reinforce the need to include realistic microlensing simulations in lensed SN studies in order to project accurate time delay precision results.

Next we attempt to find a relationship between the time delay precision and the parameters that define each microlensing magnification map ($\kappa,\gamma,s$), and therefore each magnification curve (see Section \ref{sub:micro}). Each time delay measurement is derived from exactly two images in the system, and therefore systems with more than two images will contribute more than one time delay measurement to this analysis.

We follow \citet{vernardos_joint_2018} in applying the effective convergence and shear tranformation \citep{paczynski_gravitational_1986}, which reduces the three macromodel parameters $\kappa,\gamma,s$ to just the ``effective convergence'' ($\kappa_{eff}$) and ``effective shear'' ($\gamma_{eff}$), defined by:
\begin{equation}
    \label{eq:effective}
    \kappa_{eff}=\frac{(1-s)\kappa}{1-s\kappa}, \ \gamma_{eff}=\frac{\gamma}{1-s\kappa}
\end{equation}
There is no single set of ($\kappa,\gamma,s$) parameters for any given multiply-imaged system, since each image of the SN is propagated through a distinct microcaustic field. In the comparisons that follow, we match each time delay measurement between a pair of images to the maximum $\kappa_{eff}$ and $\gamma_{eff}$ for that pair.

\begin{figure*}[ht!]
    \centering
    \includegraphics[width=.7\textwidth]{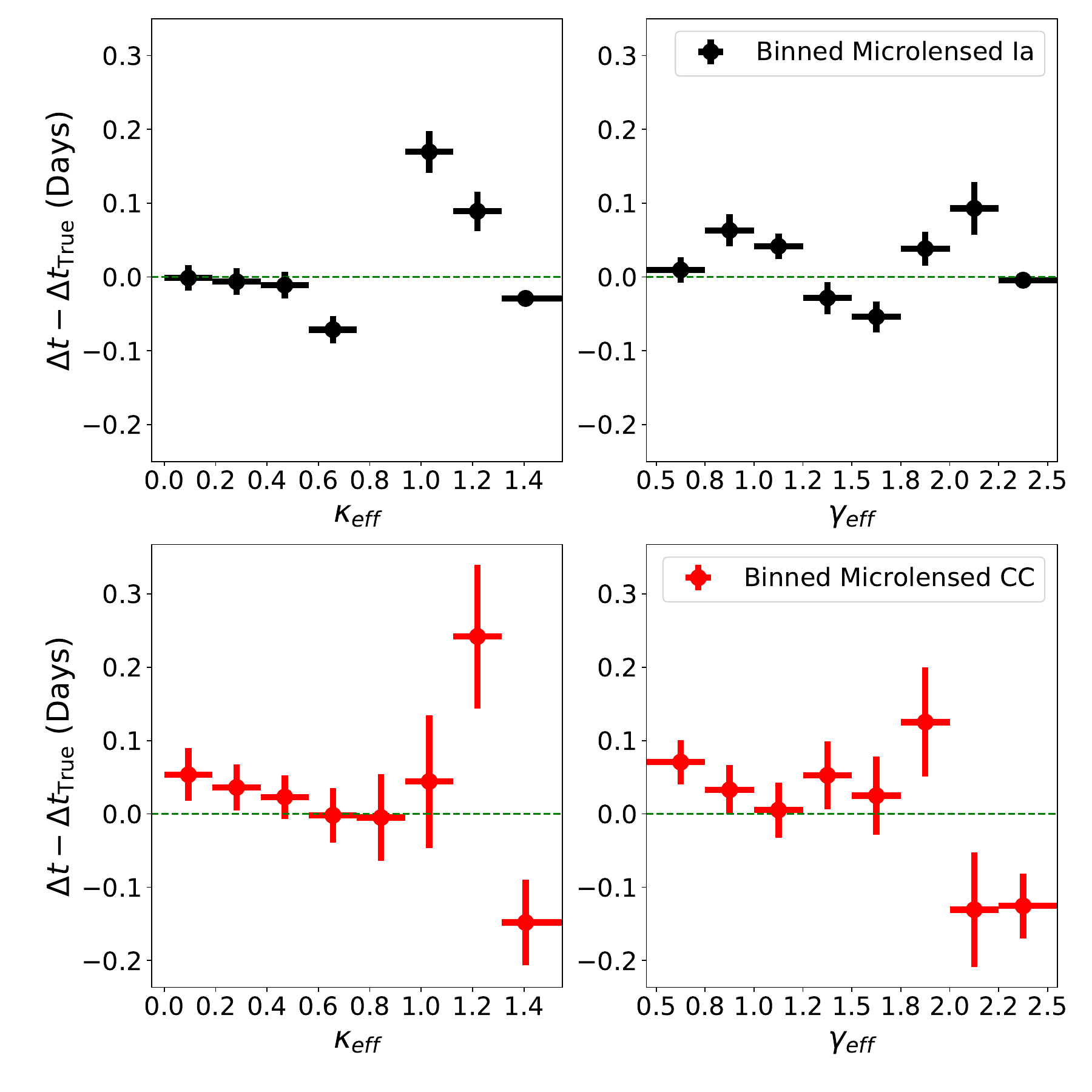}
    \caption{ Effective convergence and shear, $\kappa_{eff}$ and $\gamma_{eff}$, vs. the measurement bias for the time delay of each system, $\Delta t-\Delta t_{\rm{True}}$ in days. The black points with error bars (Top) represent binned SNIa time delay measurements, while the red points  with error bars (Bottom) represent bins that combine all CCSN sub-types.}
    \label{fig:prec_eff}
\end{figure*}

Figure \ref{fig:prec_eff} displays the measurement bias of the time delay as a function of $\kappa_{eff}$ and $\gamma_{eff}$. Higher values of $\kappa_{eff}$ and $\gamma_{eff}$ represent more extreme effects due to compact microlenses. We see that most ``normal'' values of $\kappa_{eff}$ and $\gamma_{eff}$ (see Figure \ref{fig:micro_overview}) correspond to relatively unbiased time delay measurements. Measurement accuracy appears to degrade in cases of extreme microlensing for both SNIa and CCSN, particularly for high values of $\kappa_{eff}$. 

\subsubsection{Accounting for Microlensing}
\label{sub:micro_identify}
We use the Gaussian Process Regression (GPR) method inside of {\fontfamily{qcr}\selectfont{SNTD}} to investigate our sensitivity to account for microlensing, without removing it from the light curve explicitly. This process fits each light curve of a lensed SN, and runs the GPR on the residuals to create a set of N (typically 100-1000) plausible microlensing curves that are consistent with the fitting residuals. Each possible microlensing curve is applied to the SN model and the data are refit N times, creating a distribution of time delay measurements based on the set of microlensing curves. The additional uncertainty in the time delay measurement due to microlensing combines the Gaussian $\sigma$ of this distribution and the accuracy bias relative to the non-GPR fit (see \citetalias{pierel_turning_2019} for more details).

As this process is computationally expensive, we run the GPR on a random sample of 1000 simulated SNIa with 1000 GPR samples each, and select those with a reduced-$\chi^2\leq1$ (\citetalias{pierel_turning_2019}). The issue being examined is that microlensing may cause time delay measurements for which a fitting metric like a reduced-$\chi^2$ suggests that the fit to the data is quite good, yet the measurement uncertainty is not well-characterized by the model uncertainty alone as there is a significant bias due to microlensing. We define distributions that quantify this scenario as:
\begin{equation}
\label{eq:micro_unc}
n_i=\frac{\Delta t_i-\Delta t_{\rm{i,True}}}{\sigma_i}
\end{equation}
where here $\Delta t_i$ is the $i^{th}$ measured time delay, $\Delta t_{i,\rm{True}}$ is the true $i^{th}$ time delay, $n_i$ is the number of $\sigma$'s between $\Delta t_i$ and $\Delta t_{i,\rm{True}}$, and $i$ runs over all the lensed SN in the sample. The uncertainty $\sigma_i$ is the sum in quadrature of the model uncertainty and microlensing uncertainty when running the GPR, otherwise it is simply the model uncertainty. The distributions ($n$) resulting from these two choices of $\sigma$ will be referred to as the ``GPR'' and ``model'' distributions, respectively. 

We also evaluate Equation \ref{eq:micro_unc} for the sample of non-microlensed light curves fit in Section \ref{sub:micro_precision}. This gives us a ``target distribution'', for which the model  uncertainty should predict the scatter in time delay measurements because there is no microlensing causing a bias. We compare the GPR and model distributions to the target distribution by way of a root mean square (RMS) comparison, and with a Kolmogorov-Smirnov (KS) test (Tables \ref{tab:micro_rms}-\ref{tab:micro_ks}). For the RMS comparison (Table \ref{tab:micro_rms}) the ``Model-No Microlensing'' distribution approaches the expected result of RMS$\simeq1$, while the ``Model-With Microlensing'' distribution is significantly higher suggesting an underestimation of the uncertainties associated with microlensing. The GPR result brings this back below RMS$=1.5$, an improvement over the model uncertainties alone. 

Similarly, we compare the GPR and model distributions to the target distribution, and find that the model distribution cannot be said to match the target distribution (P<0.01) while the GPR distribution once again matches the target distribution (P>0.01). For both the RMS comparison and the KS-test, we conclude that model error alone is insufficient to describe the uncertainty introduced by microlensing, but that the GPR is able to more accurately characterize the full uncertainty.  For cases where microlensing has a significant impact on the light curve, as in these simulations, the microlensing uncertainty produced by the GPR is required to fully characterize the time delay uncertainty.

\begin{table}[h]
\centering

\caption{\label{tab:micro_rms} Root mean square (RMS) for equation \ref{eq:micro_unc} for the three separate distributions: model uncertainty with no microlensing, model uncertainty with microlensing, and the GPR fitting with microlensing. }
\begin{tabular*}{\columnwidth}{@{\extracolsep{\stretch{1}}}*2{c}}
\toprule
  \multicolumn{1}{c}{\textbf{Distribution}}&\multicolumn{1}{c}{\textbf{RMS}}\\

\hline
Model-No Microlensing&1.1\\
Model-With Microlensing&1.6\\
GPR&1.4\\

\end{tabular*}

\end{table}

\begin{table}[t]
\centering
\caption{\label{tab:micro_ks} Result of a KS test that compares the model and GPR distributions to the target distribution, which comprises time delay measurements of non-microlensed light curves. }
\begin{tabular}{ccrr}
\toprule
 \multicolumn{1}{c}{\textbf{Distribution}}&\multicolumn{1}{c}{$\mathbf{\sigma}$} &\multicolumn{1}{c}{\textbf{KS Statistic}}&\multicolumn{1}{c}{\textbf{P-Value}} \\

\hline
Model&$\sigma_{\rm{Model}}$&0.090&0.004\\
GPR&$\sqrt{\sigma_{\rm{Model}}^2+\sigma_\mu^2}$&0.058&0.153\\

\end{tabular}

\end{table}

\subsubsection{Type Ia Standardization}
\label{sub:micro_ia}
The process of SNIa standardization first requires some selection cuts based on light curve fitting (SNIa with extreme light curve properties are not reliably standardizable). For SNIa that pass the cuts, an empirical model can be fit to the observed light curve to measure a corrected peak magnitude.

In this section we investigate how many observed lensed SNIa from the \textit{Roman Space Telescope} would pass typical selection cuts, and then explore the standardizablity of those selected SNIa. Each SNIa light curve is fit with the extended SALT2 model \citep{guy_salt2:_2007,pierel_extending_2018}, which has 4 free parameters: An overall brightness scale parameter ($x_0$), a light curve ``stretch'' parameter ($x_1$), a ``color'' parameter ($c$), and the time of peak ($t_0$). In \citet{betoule_improved_2014}, the relevant quality cuts required that the fitted $x_1,c$ parameters satisfy $|x_1|<3,|c|<0.3$ and the uncertainties on the $x_1$ and $t_0$ parameters satisfy $\sigma(x_1)<1, \ \sigma(t_0)<2$. In that analysis, $\sim79\%$ of the SNIa population passed these cuts. 

\begin{figure}[h!]
    \centering
    \includegraphics[trim={0cm 0cm 1cm 3.5cm},clip,width=.5\textwidth]{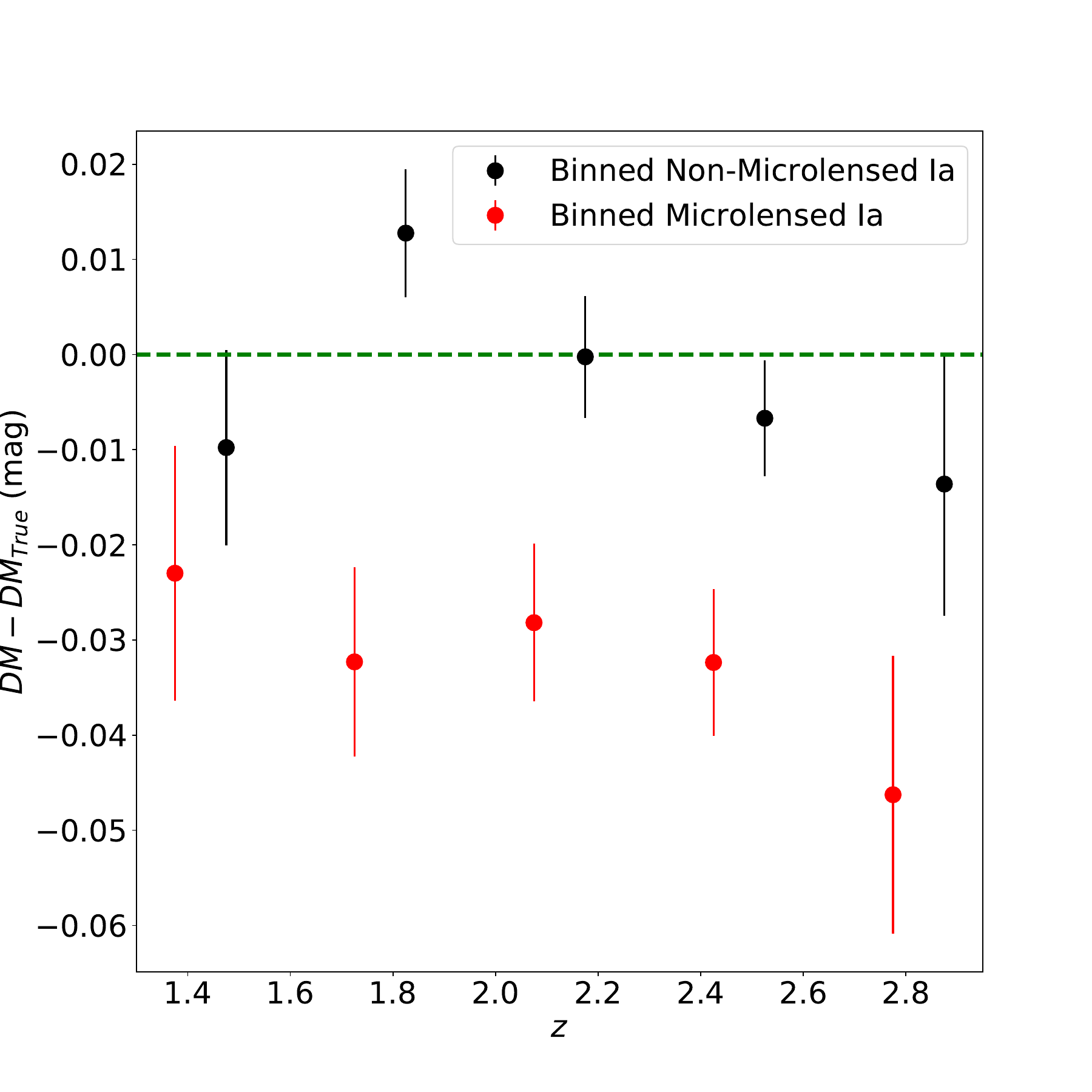}
    \caption{The measured SNIa distance modulus (DM) minus the true DM as a function of redshift. Black points with error bars are binned simulated light curves without microlensing applied, while the red points with error bars have microlensing applied. The green dashed line represents zero bias in the DM measurement.}
    \label{fig:dismod_z}
\end{figure}

We use the same sample from the previous Section \ref{sub:micro_precision} to determine the difference between macro-lensed and macro+micro-lensed SNIa standardization. We find that 77\% of non-microlensed and 72\% of microlensed light curves pass all quality cuts. We next assume static nuisance parameters in the SNIa distance estimator formula \citep[$\alpha,\beta,M_B$;][]{tripp_two-parameter_1998} and calculate the distance modulus (DM) accuracy as a function of redshift ($DM_i(z_i)-DM_{i,\rm{True}(z_i)}$; Figure \ref{fig:dismod_z}). The corresponding global biases are presented in Table \ref{tab:selection_cuts}.  We find a small bias ($\sim$-0.01 mag) in DM measurements for non-microlensed light curves, and a more substantial bias introduced by microlensing ($\sim-0.03$ mag). It is plausible that this bias could be corrected with large-scale simulations similar to those currently employed to remove selection biases for SNIa cosmology \citep[e.g.,][]{scolnic_measuring_2016,kessler_correcting_2017}, as this bias is roughly half of the correction introduced to account for the correlation of Hubble residuals with SN host galaxy mass \citep{jones_should_2018}. 
\begin{table*}[t!]
\centering
\caption{\label{tab:selection_cuts} The result of calculating the distance modulus (DM) for light curve simulations with and without microlensing applied. }
\begin{tabular*}{\textwidth}{@{\extracolsep{\stretch{1}}}*4{r}}
\toprule
 \multicolumn{1}{c}{\textbf{Sample}}&\multicolumn{1}{c}{\textbf{Passing Cuts}}& \multicolumn{1}{c}{\textbf{DM Bias}} &\multicolumn{1}{c}{\textbf{DM Precision}} \\
 &&\multicolumn{1}{c}{mag}&\multicolumn{1}{c}{mag}\\
\hline
No Microlensing&77\%&$-0.007\pm0.0030$&0.2\\
Microlensing&73\%&$-0.030\pm0.0002$&0.2\\
\end{tabular*}

\end{table*}
As all microlensing in this work is achromatic there is essentially zero bias introduced by lensing into the measurement of $c$, meaning that all bias in DM is essentially from inaccurate $x_1$ measurements. It is likely that measurements of $c$ would also lose accuracy with the implementation of chromatic microlensing, but the exact impact on standardizability should be investigated by future work.

\section{Cosmological Constraints}
\label{sec:cosmo}
The lensed SN yields from Section \ref{sec:NGRST} and the measured time delay precision for SNIa and CCSN from Section \ref{sec:delays} are now propagated through to expected constraints on cosmology. We follow the methodology of \citet{coe_cosmological_2009}\footnote{Now implemented as the cosmology module in v2 of {\fontfamily{qcr}\selectfont{SNTD}}} to estimate how each survey strategy constrains parameters in a given cosmological model. The basis of these constraints is from a ratio of angular diameter distances that appear in the time delay equation:
\begin{equation}
    \label{eq:td}
    \Delta t=\frac{(1+z_L)}{c}\frac{D_LD_S}{D_{LS}}\Big[\frac{1}{2}|\vec\theta-\vec\beta|^2-\phi\Big].
\end{equation}
Where ``L'' is for the Lens and ``S'' is for the source, so that $z_L$ is the redshift of the lens and $D_L$, $D_S$, $D_{LS}$ are the angular diameter distances to the lens, to the source, and from the lens to the source respectively. The first term in brackets contains the source image positions ($\vec\theta$) and the true (unlensed) source  position ($\vec\beta$), while the second term is the projected lens potential $\phi$. These two terms comprise the lens model component of the time delay equation, which allows us to split up equation \ref{eq:td} into a cosmological component, \TC, and a lensing component, \TL:
\begin{equation}
    \label{eq:td2}
    \Delta t=\TC\,\TL
\end{equation}
where:
\begin{equation}
    \label{eq:tc_tl}
    \TC\equiv\frac{(1+z_L)}{c}\frac{D_LD_S}{D_{LS}};  \ \ \ \  \TL\equiv\Big[\frac{1}{2}|\vec\theta-\vec\beta|^2-\phi\Big].
\end{equation}
Therefore with a measured time delay $\Delta t$ and a well-constrained lens model to estimate \TL, we can derive a constraint on \TC, which holds the cosmological dependencies. For the remainder of this work we set the estimated lens model uncertainty as $\dTL=5\%$, which is reasonable based on previous work \citep{coe_cosmological_2009,suyu_dissecting_2010,shajib_strides_2020}, and set the value of \TC~separately for SNIa and CCSN based on our \textit{Roman} simulations. All of our cosmological constraints contain statistical errors only \citep{coe_cosmological_2009}, but we discuss potential sources of systematic uncertainties and their impacts for these projections in Section \ref{sub:systematics}.

Now we turn to the \TC~term in equation \ref{eq:tc_tl}. Each angular diameter distance in the ratio is a function of the normalized Hubble Parameter $E(z)$:
\begin{equation}
    \label{eq:hubble}
    E(z)=\frac{H(z)}{H_0}=\sqrt{\Omega_m(1+z)^3+\Omega_k(1+z)^2+\Omega_\Lambda}
\end{equation}
where $H(z)$ is known as the Hubble parameter, $H_0$ the famous Hubble-LeMa\^itre constant, and $\Omega_m,\Omega_k,\Omega_\Lambda$ are the matter density, curvature, and dark energy density terms respectively. 

Now we proceed to specific cosmological models, and in this work we choose two separate models to investigate. In both models we assume flatness, which is to say that the curvature term $\Omega_k=0$ (so that $\Omega_m+\Omega_\Lambda=1$) and the angular diameter distance can be written as:
\begin{equation}
    \label{eq:add}
    D_A(z_1,z_2)=\frac{c}{H_0(1+z_2)}\int\limits_{z_1}^{z_2}\frac{dz\prime}{E(z\prime)}
\end{equation}
The dependence of the term \TC~on $H_0$ is clear from Equation \ref{eq:add}, and the dependence on the density parameters $\Omega_m$ and $\Omega_\Lambda$ is clear from Equation \ref{eq:hubble}. For convenience, we define a term $E_A$ that takes the from of the integral in Equation \ref{eq:add}:
\begin{equation}
    \label{eq:e_a}
    E_A(z_1,z_2)\equiv\int\limits_{z_1}^{z_2}\frac{dz\prime}{E(z\prime)}
\end{equation}
Combining the above equations, we summarize these dependencies as:
\begin{equation}
    \label{eq:tc_final}
    \TC=\frac{\Sigma(\Omega_m,\Omega_k,\Omega_\Lambda)}{H_0}; \ \ \Sigma\equiv\frac{E_LE_S}{E_{LS}}
\end{equation}
Where $E_L,E_S,E_{LS}$ are $E_A(0,z_L),E_A(0,z_S),E_A(z_L,z_S)$, respectively. The dependence of \TC~on each of the cosmological parameters can now be sampled for a given set of lens and source redshift combinations, as well as time delay precision $\delta(\Delta t)$, which propagates into the cosmological precision, \dTC. In the following sections, we perform this sampling with each of the mock survey catalogs produced in Section \ref{sec:NGRST} and the time delay precision measured in Section \ref{sec:delays}, translated into a percent precision for each catalog entry. 

For each cosmological model, we set the ``true'' values of these parameters to $H_0=70 \ \rm{km} \ \rm{s}^{-1} \ \rm{Mpc}^{-1}$, $\Omega_m=0.3$, $\Omega_\Lambda=0.7$. In addition to the assumption of flatness, each model we investigate incorporates so-called Cold Dark Matter \citep[CDM][]{aghanim_planck_2018}. Finally we assume perfect knowledge, or effectively negligible uncertainty on, each lens and source redshift. 

To perform the analysis, we use Fisher matrices. Fisher matrices do not retain the full cosmological information present in Equation \ref{eq:hubble} and approximate all uncertainties as Gaussians, but are usually sufficient if cosmological parameters are constrained close to their true values \citep{coe_cosmological_2009}. The elements of a Fisher matrix can be defined in the following way:
\begin{equation}
    \label{eq:fisher}
    F_{ij}=\frac{1}{2}\frac{\partial\chi^2}{\partial p_i\partial p_j}
\end{equation}
where here the $\chi^2$ is the goodness of fit metric between an assumed cosmology and the true cosmology, characterized by the formalism of the previous section. The shape and orientation of each contour defined by a Fisher matrix is weakly dependent on the lens and source redshift ratios, and the size is driven by a combination of the number of observed lensed SN and the relative precision of each measured time delay (i.e. the length of the time delay itself).

\

\subsection{Flat wCDM}
\label{sub:wcdm}

The first of the two models we investigate is the Flat$w$CDM model, whereby the dark energy density term $\Omega_\Lambda$ is varied to satisfy a dark energy equation of state with $p=w\rho$:
\begin{equation}
    \label{eq:wcdm}
    E(z)=\sqrt{...+\Omega_\Lambda(1+z)^{3(1+w)}}
\end{equation}
Where the ellipsis represents unchanged terms compared to equation \ref{eq:hubble}. Here we set the true value of $w$ to be -1, corresponding to a universe with a cosmological constant. The addition to Equation \ref{eq:hubble} causes an extra dependence for \TC~compared to Equation \ref{eq:tc_final}:
\begin{equation}
    \label{eq:tc_w}
    \TC=\frac{\Sigma(\Omega_m,\Omega_k,\Omega_\Lambda,w)}{H_0}.
\end{equation}

\begin{figure*}[h!]
    \centering
    \includegraphics[trim={1.3cm 1cm .7cm 0cm},clip,width=\textwidth]{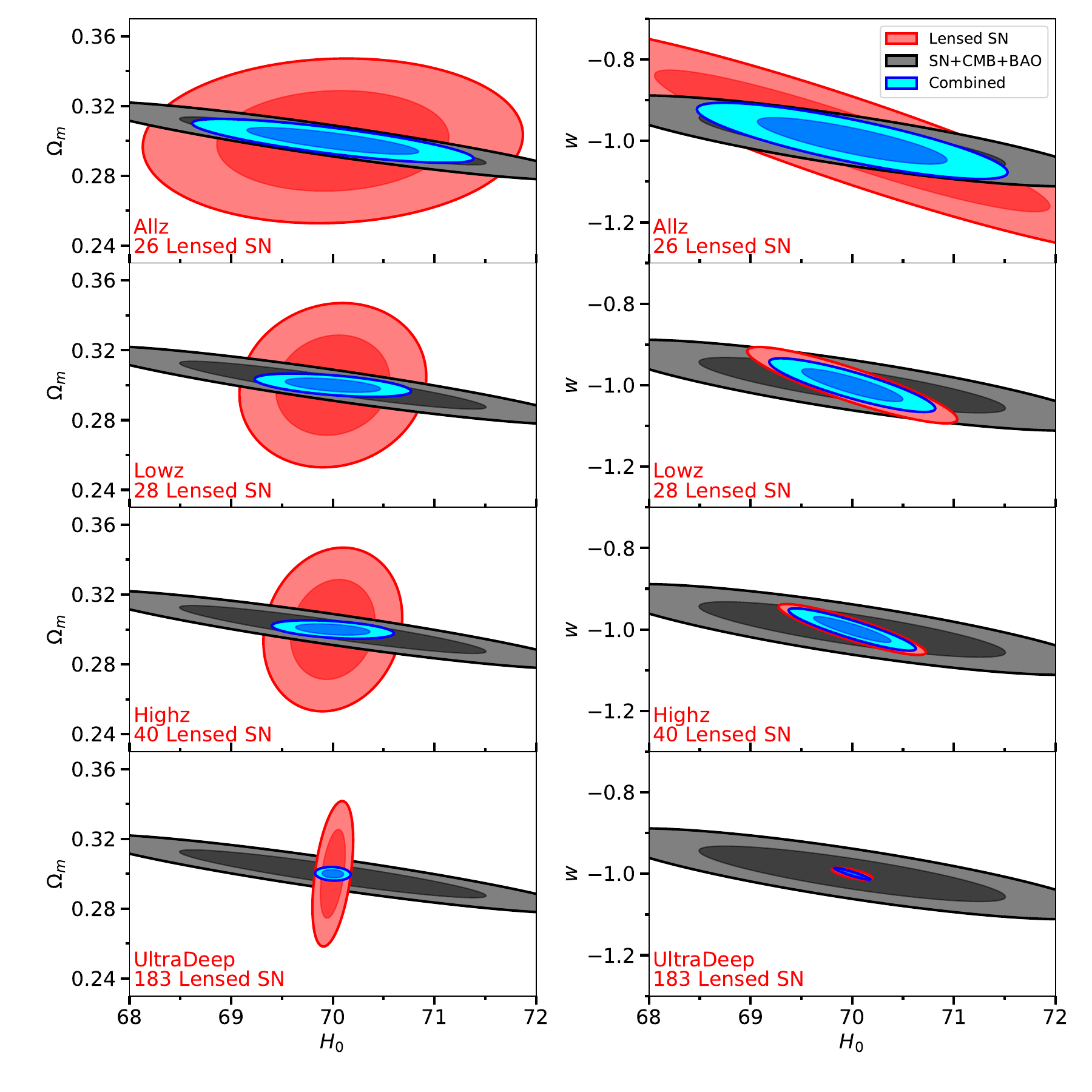}
    
    \caption{Result of the Fisher matrix analysis in this section, for the ($H_0,\Omega_m$) (left column) and ($H_0,w$) (right column) parameters in the Flat$w$CDM cosmology. Each row corresponds to a survey strategy, shown in red text. The red contours represent the (1 \& 2$\sigma$) constraints obtainable from lensed SN time delays alone, with a prior from \citet{scolnic_complete_2018} on each parameter. The black contours show the current constraints from the Fisher matrix representing the CMB, BAO, and SNIa measurements of these parameters. The blue contours show the combined constraints, which are quoted in Table \ref{tab:wcdm_cosmo}.}
    \label{fig:wcdm_fisher}
\end{figure*}

We now investigate the constraints on cosmological parameters for each SNIa survey strategy discussed in Section \ref{sec:NGRST}, using the Fisher matrix analysis described above. We examine what a \textit{Roman Space Telescope} lensed SN sample would contribute when combined with the results of present day constraints from the Cosmic Microwave Background \citep[CMB][]{planck_collaboration_planck_2016}, Baryon Acoustic Oscillations \citep[BAO]{anderson_clustering_2014}, and SNIa distance measurements \citep{scolnic_complete_2018}. 
We use Fisher matrices derived from the full CosmoMC \citep{lewis_cosmological_2002} chains present in \citet{scolnic_complete_2018} to make this comparison. More precisely, these Fisher matrices are inverses of each probe's respective covariance matrix, calculated from each full CosmoMC chain. 
The results of the Fisher matrix analysis are summarized in Table \ref{tab:wcdm_cosmo}, with cosmological parameter relationships for $(H_0,\Omega_m)$ and $(H_0,w)$ shown for each survey in figure \ref{fig:wcdm_fisher}. The results shown in Table \ref{tab:wcdm_cosmo} are based upon the combination of current probes, and the expected sample of lensed SN for each strategy. Every value is the mean of 10 distinct analyses, each with a random sample of lensed SN from the catalogs in Section \ref{sec:NGRST}. The current precision for each parameter is given for comparison, measured by combining constraints from the CMB, BAO, and SNIa alone \citep{scolnic_complete_2018}. While of course the addition of any lensed SN sample improves constraints on all cosmological parameters, we note that the Highz strategy provides the most significant gains in precision. As we have noted throughout this paper, the UltraDeep strategy is not a realistic proposal for a SNIa survey, and should be considered as the upper limit on what is possible from lensed SN from \textit{Roman}.

Note that we have made a conscious choice to use current, published cosmological constraints for this comparison rather than using projections of future probes, such as the large SNIa and weak lensing measurements that will come from the Rubin Observatory, the Euclid mission, and \textit{Roman} itself.  This choice obviates the need to adjudicate the validity and compatibility of those other studies.  A comparison to projections of future cosmological constraints would certainly be of interest, but it is beyond the scope of this work. 
To enable such investigations in the future, our Fisher matrices have been included in the cosmology module inside of the SNTD package, which was used to perform the present analysis. This package, or simply the Fisher matrices provided for each survey strategy, can be combined with assumptions about future probes to determine contributions from lensed SN cosmology.

\begin{table}[t!]
\centering
\caption{\label{tab:wcdm_cosmo} Estimated final uncertainty on each parameter in the Flat$w$CDM cosmology from our Fisher matrix analysis. }
\begin{tabular*}{\columnwidth}{@{\extracolsep{\stretch{1}}}*4{r}}
\toprule

\multicolumn{1}{c}{\textbf{Probes}} & \multicolumn{1}{c}{$\mathbf{\delta w}$}&\multicolumn{1}{c}{$\mathbf{\delta\Omega_m}$}&\multicolumn{1}{c}{$\mathbf{\delta H_0}$}\\

CMB+BAO+SNIa&$0.04$&$0.008$&0.86\\
\hline
 ...+Lowz Lensed SN&0.03&0.003&0.32\\
...+Allz Lensed SN&0.04&0.005&0.56\\
...+Highz Lensed SN&0.02&0.002&0.27\\
...+UltraDeep Lensed SN&0.01&0.002&0.07\\
\end{tabular*}

\end{table}

\subsection{Flat $w_0w_a$CDM}
\label{sub:w0wacmd}
\begin{figure*}[h!]
    \centering
    \includegraphics[trim={1.3cm 1cm .7cm 0cm},clip,width=\textwidth]{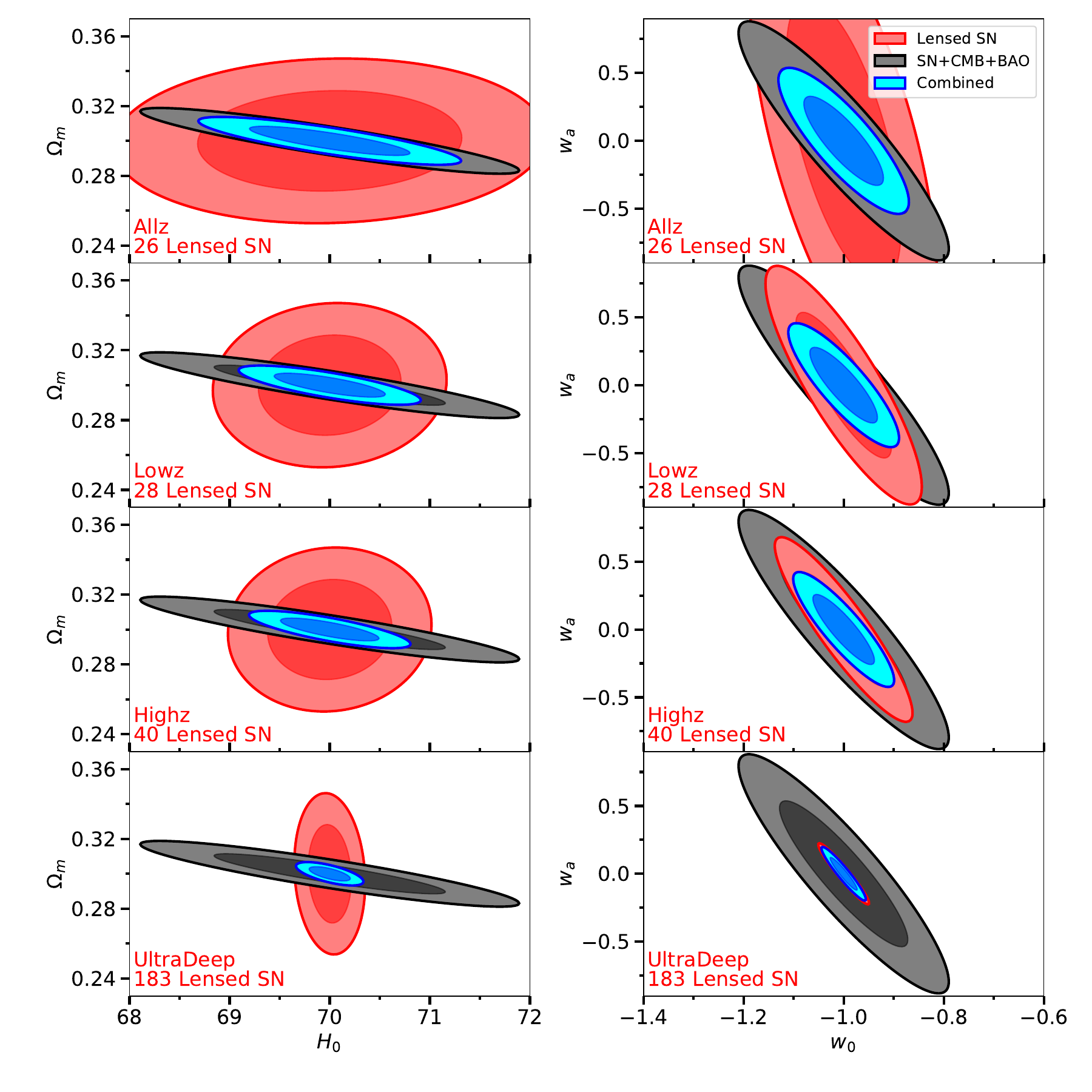}
    \caption{Result of the Fisher matrix analysis in this section, for the ($H_0,\Omega_m$) (left column) and ($w_0,w_a$) (right column) parameters in the Flat$w_0w_a$CDM cosmology. Each row corresponds to a survey strategy, shown in red text. The red contours represent the (1 \& 2$\sigma$) constraints obtainable from lensed SN time delays alone, with a prior from \citet{scolnic_complete_2018} on each parameter. The black contours show the current constraints from the Fisher matrix representing the CMB, BAO, and SNIa measurements of these parameters. The blue contours show the combined constraints, which are quoted in Table \ref{tab:wcdm_cosmo}.}
    \label{fig:w0wa}
\end{figure*}

The second cosmological model we investigate is the Flat$w_0w_a$CDM model \citep{chevallier_accelerating_2001}:
\begin{equation}
    \label{eq:w0wacdm}
    E(z)=\sqrt{...+\Omega_\Lambda(1+z)^{3(1+w_0+w_a)}{\rm{exp}}\Big(\frac{-3w_az}{1+z}\Big)}
\end{equation}
So that equation \ref{eq:tc_final} becomes:
\begin{equation}
    \label{eq:tc_w0wa}
    \TC=\frac{\Sigma(\Omega_m,\Omega_k,\Omega_\Lambda,w_0,w_a)}{H_0}
\end{equation}
Where $w_0,w_a$ correspond to an evolving dark energy equation of state defined by:
\begin{equation}
    \label{eq:w0wa_eos}
    p=\Big[w_0+w_a\Big(\frac{z}{1+z}\Big)\Big]\rho
\end{equation}
Here we set the true values of $w_0$ and $w_a$ to be -1 and 0, respectively, which as in the previous section corresponds to a universe with a cosmological constant. The results of the Fisher matrix analysis are summarized in Table \ref{tab:w0wacdm_cosmo}, with cosmological parameter relationships for ($H_0,\Omega_m)$ and ($w_0,w_a$) shown for each survey strategy in figure \ref{fig:w0wa} ). The results shown in Table \ref{tab:w0wacdm_cosmo} are based upon the combination of current (or future) probes, and the expected sample of lensed SN for each strategy.  Every value is the mean of 10 analyses, each with a random sample of lensed SN from the catalogs in Section \ref{sec:NGRST}. 

\begin{figure*}[ht!]
\centering
\includegraphics[ width=\textwidth]{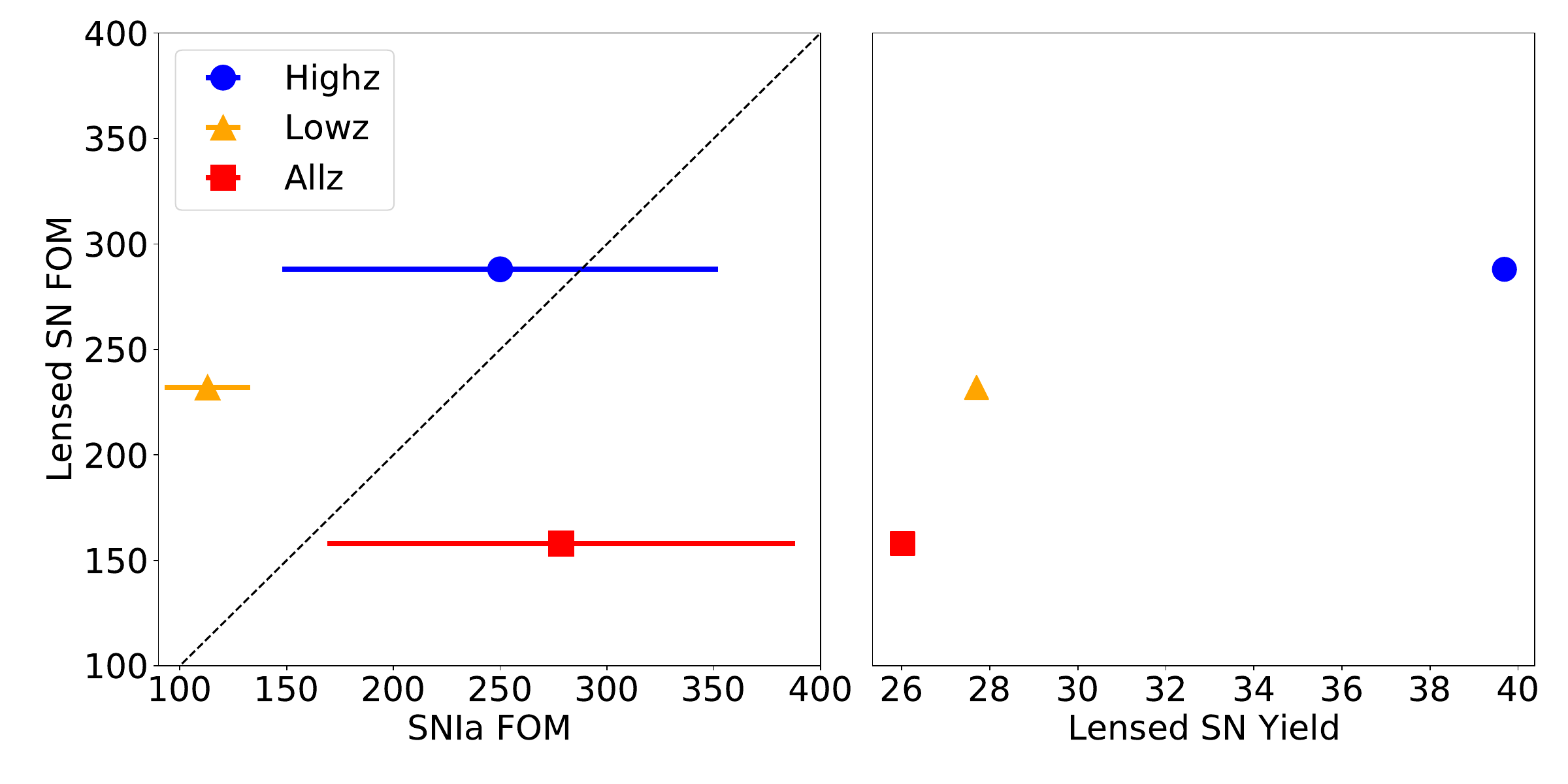}
\caption{\label{fig:Ia_lensing_fom} Comparing each SNIa survey strategy using FOM and lensed SN yield. In each panel, the y-dimension shows the FOM of the Lowz (red square), Allz (orange triangle), and Highz (blue circle) strategy, combined with current probes (Table \ref{tab:w0wacdm_cosmo}, Figure \ref{fig:w0wa}). In the left panel, the x-dimension shows the FOM projections for each strategy from \citetalias{hounsell_simulations_2018}, with the error bars denoting a range in possible systematics. We note that for lensed SN, increasing the sample will actually improve systematics because each lensing system is unique, giving us new information with each discovery. The dashed line represents the line of equivalence between SNIa FOM and lensed SN FOM. In the right panel, the abscissa shows the expected lensed SN yields from Table \ref{tab:rst_survey}. Our conclusion is that, in general, the Highz SNIa strategy will yield the greatest return for a combination of SNIa and lensed SN cosmology.} 
\end{figure*}

We use the Figure of Merit (FOM), defined as the area of the $95^{\rm{th}}$ percentile contour in ($w_0,w_a$) space \citep{albrecht_report_2006}, to compare the different SNIa survey strategies. The FOM for each strategy are reported in Table \ref{tab:w0wacdm_cosmo}, as well as the current FOM and precision using combined constraints from the CMB, BAO, and SNIa alone for comparison \citep{scolnic_complete_2018}. Figure \ref{fig:Ia_lensing_fom} compares strategies based on lensed SN FOM, SNIa FOM, and lensed SN yield. We find that considering all of these aspects, the Highz survey is the best overall strategy when considering SNIa and lensed SN cosmology.

\

\begin{table}[h]
\centering
\caption{\label{tab:w0wacdm_cosmo} Estimated final uncertainty on each parameter in the Flat$w_0w_a$CDM cosmology from our Fisher matrix analysis. }
\begin{tabular*}{\columnwidth}{@{\extracolsep{\stretch{1}}}*6{r}}
\toprule

\multicolumn{1}{c}{\textbf{Probes}} & \multicolumn{1}{c}{$\mathbf{\delta w_0}$}& \multicolumn{1}{c}{$\mathbf{\delta w_a}$}&\multicolumn{1}{c}{$\mathbf{\delta\Omega_m}$}&\multicolumn{1}{c}{$\mathbf{\delta H_0}$}&\multicolumn{1}{c}{FOM} \\

CMB+BAO+SNIa&$0.09$&0.38&$0.008$&0.86&77\\
\hline
 ...+Lowz lensed SN&0.04&0.18&0.005&0.38&232\\
...+Allz lensed SN&0.05&0.22&0.006&0.53&158\\
...+Highz lensed SN&0.04&0.17&0.004&0.35&288\\
...+UltraDeep lensed SN&0.02&0.08&0.003&0.14&2162\\
\end{tabular*}

\end{table}

\subsection{Systematics}
\label{sub:systematics}
As discussed at the beginning of this section, all projections in this work assume statistical uncertainties alone. It is possible that the impacts from chromatic microlensing or redshift-dependent selection effects could represent ignored systematic uncertainties for $\TC$. We cannot yet investigate the systematic uncertainty associated with chromatic microlensing across all SN types due to a lack of available models (see Section \ref{sub:micro}), and a selection bias is expected to introduce a small and correctable uncertainty \citep[e.g.,][]{collett_observational_2016}.

However, we are ignoring potentially significant sources of systematic uncertainty in our choice of $\TL$. First, previous work has found that astrometric uncertainties can become dominant in the cosmological error budget if a relative astrometric precision of $\lesssim1-10$ mas cannot be achieved \citep{birrer_astrometric_2019}. While this should be considered for surveys such as the LSST, astrometric uncertainties are expected to be sub-dominant for \textit{Roman Space Telescope} observations where a precision of 1 mas or better is predicted \citep{sanderson_astrometry_2019}. Second, there are known parameter degeneracies in lens modeling that previous work suggests could contribute a systematic error term on the order of $\mathbf{\gtrsim10\%}$ \citep{coe_cosmological_2009,kochanek_overconstrained_2020}. While it is believed that the addition of high-quality velocity dispersion measurements and improvements such as a hierarchical Bayesian lens modeling method could drastically reduce this systematic effect for future discoveries \citep{birrer_mass-sheet_2016,birrer_tdcosmo_2020,birrer_tdcosmo-strategies_2020}, this uncertainty term is unlikely to disappear entirely.  Additionally, while the standardizability of lensed SNIa is still an open question (see Section \ref{sub:micro_ia}), it is likely that each discovery of a lensed SNIa will provide some unique constraint for lens models that will further limit the impact of these potential biases \citep[e.g.,][]{foxley-marrable_impact_2018}.

To estimate the impact that such systematic biases could have on the FOM, we repeat the Fisher matrix analysis from Section \ref{sub:w0wacmd} with an additional systematic uncertainty.
In truth these terms should be represented by covariant biases among all discovered lenses, as opposed to a simple systematic uncertainty term, which would propagate differently through to cosmology. 
However, to get an estimate of the impact for cosmology in an extreme case, we adopt a systematic uncertainty of 10\%.
Even in this highly conservative scenario, the FOM still improves by $60-70\%$ relative to the current value. Also, the Highz strategy maintains its place as the most valuable for lensed SN cosmology. Future work should investigate in detail the manner in which these covariant biases are truly propagated through to cosmology, as well as the potential impacts of kinematics or SNIa standard candle constraints.

\section{Discussion and Conclusions}
\label{sec:discussion}
In this work, we have created mock catalogs of lensing systems that match characteristics of those we expect to observe with the \textit{Roman Space Telescope}. These catalogs will enable many future investigations into lensed SN and the \textit{Roman Space Telescope}, as they cover a range of anticipated surveys including 3 SNIa survey strategies, a version of the HLS, and a hypothetical survey optimized for finding lensed SN\footnote{All of our data products are available at Mikulski Archive for Space Telescopes (MAST) via \dataset[10.17909/t9-k8w7-zk32]{\doi{10.17909/t9-k8w7-zk32}}.}. 

We next simulated a large sample of lensed SN light curves, including microlensing effects, and fit each light curve to get an estimate of the time delay precision obtainable with \textit{Roman} for each SN type. Various tests related to the impact of microlensing were conducted. We projected these precision results across each SNIa survey strategy, then combined the precision results with the yields from Section \ref{sec:NGRST} to estimate cosmological constraints obtainable for each \textit{Roman} SNIa survey strategy. We find that a survey optimized for high-redshift SNIa discovery (the ``Highz'' strategy of \citetalias{hounsell_simulations_2018}) would be the most effective for lensed SN time delay cosmography. It is possible that such a strategy could be sub-optimal if there is an un-accounted for redshift dependent systematic bias in time delay measurements (e.g. from chromatic microlensing) or lens modeling methods, which should be investigated in future work.

The results of the work described above can be summarized as follows:
\begin{enumerate}
    \item \textbf{Lensed SN Rates}--For each catalog, we have determined the expected lensed SN yield for both SNIa and CCSN, resulting in a median expectation of $\sim11$ SNIa and $\sim20$ CCSN from the various SNIa survey strategies. Although the HLS yield does not include light curves and would require a follow-up campaign for each lensed SN, we still project that it will discover $\sim200$ lensed SN, including $\sim60$ SNIa and  $\sim224$ CCSN. Finally, the survey strategy ``UltraDeep'' investigated in this work would discover $\sim56$ SNIa.
    \item \textbf{Time Delay Precision}--We find that subject to our simplifying assumptions, a precision of $\sim2$ days for SNIa and $\sim3$ days for CCSN can be expected, with small but statistically significant measurement biases of $\sim0.01-0.04$ days for CCSN only ($<1\%$ for 80\% of the simulated SN). The SNIa survey strategies should therefore observe $\sim$5, 2, 10 SN at $<5\%$ precision for Lowz, Allz, Highz respectively.
    \item \textbf{Microlensing Impact on Time Delays}--We find that low (and more likely) values of effective convergence ($\kappa_{eff}\lesssim1$) result in a small accuracy bias, while high values cause time delay accuracy to degrade by $\sim0.2-0.3$ days for both SNIa and CCSN. The time delay accuracy relationship with effective shear is not as obvious, and the biases are smaller ($\sim0.1$ days).
    \item \textbf{SNIa Standardization}--Achromatic microlensing causes a $\sim5\%$ loss in lensed SNIa that pass fiducial selection cuts. Of those that pass cuts, microlensed SN produce a biased measurement of the distance modulus ($-0.03\pm0.0002$ mag) with identical statistical precision to non-microlensed light curves. Including chromatic microlensing effects will likely degrade this further, as measuring the SALT2 ``c''  parameter accurately will be more difficult. 
    \item \textbf{Cosmological Constraints}--We find significant evidence that the Highz SNIa survey strategy is preferred when considering the FOM (in a Flat$w_0w_a$CDM cosmology) possible from both SNIa and lensed SN cosmology. Additionally, we project the Highz survey to deliver roughly double the precision of each cosmological parameter in the Flat$w$CDM cosmology compared to the Allz survey, which had the highest projected SNIa FOM. We also consider the impact of covariant biases in the lens modeling process, using the highly conservative case of simply adding large systematic uncertainties to the Fisher matrix analysis. While such systematic uncertainties would certainly degrade the precision on cosmological parameters, the core conclusions remain unchanged:  a Highz strategy is preferred, and lensed SN cosmology will be a valuable addition to the \textit{Roman} cosmology toolkit. A detailed analysis is needed to correctly propagate such covariant biases from lens modeling to cosmology to investigate their true impact. 
\end{enumerate}
 
Throughout this paper we have assumed that each lensed SN has a spectroscopic redshift assigned to it, and that each SN type is known (for IIn and IIP, we assume that each SN has been identified simply as ``Type II''). Future work should investigate how we can obtain spectroscopic redshift measurements for each lensed SN observed by \textit{Roman}. While this seems plausible and such observations could simultaneously determine the SN type, for other missions like the Rubin Observatory's LSST, this will not be the case and a study of measuring time delays without redshift and/or SN type information is important. Another simplifying assumption made here is that of achromatic microlensing, which was necessary due to a lack of 2-D projected specific intensity profiles for CCSN. Future work should improve upon this assumption when such models become available, in order to investigate possible systematic impacts on time delay measurements and further degradation in SNIa standardizability from introducing chromatic effects. 

The \textit{Roman Space Telescope} is expected to make significant strides for investigations into dark energy, particularly through SNIa cosmology. We have found that, without varying the SNIa survey strategies, \textit{Roman} will discover a large number of lensed SNIa and CCSN useful as cosmological probes. While we find the Highz strategy would be the most optimal for combined constraints from SNIa and lensed SN, it's clear that any imaging-heavy survey will enable significant contributions from SN time delay cosmography with the \textit{Roman Space Telescope}. 

\noindent \acknowledgments

\noindent We would like to thank S. Suyu, L. Moustakas, S. Birrer, and T. Treu for useful discussions and feedback, as well as the anonymous referee for helpful comments and suggestions. This  work  was  supported by 
the National Aeronautics and Space Administration (NASA)
Headquarters under the NASA Future Investigators in Earth and Space Science and Technology (FINESST) award 80NSSC19K1414, and by NASA  contract  No.  NNG17PX03C  issued  through  the \textit{Roman} Science Investigation Teams Program. 
M.O. was supported in part by World Premier International
Research Center Initiative (WPI Initiative), MEXT, Japan, and JSPS
KAKENHI Grant Number JP20H00181. T.A. acnkowledges support from Proyecto Fondecyt N 1190335 and the Ministry for the Economy, Development, and Tourism's Programa Inicativa Cient\'ifica Milenio through grant IC 12009.

\bibliographystyle{aas}

\pagebreak

\appendix
\setcounter{table}{0}
\renewcommand{\thetable}{A\arabic{table}}

\setcounter{figure}{0}
\renewcommand{\thefigure}{A\arabic{figure}}

\begin{longtable}{ccc} 
\textbf{Model Name} & \textbf{Type} & \textbf{Reference} \\
\hline
nugent-sn1a	&	SN Ia	&	\citet{nugent_k_2002}	\\
nugent-sn91t	&	SN Ia	&	\citet{stern_discovery_2004}	\\
nugent-sn91bg	&	SN Ia	&	\citet{nugent_k_2002}	\\
nugent-sn1bc	&	SN Ib/c	&	\citet{levan_grb_2005}	\\
nugent-hyper	&	SN Ib/c	&	\citet{levan_grb_2005}	\\
nugent-sn2p	&	SN IIP	&	\citet{Gilliland_high-redshift_1999}	\\
nugent-sn2l	&	SN IIL	&	\citet{Gilliland_high-redshift_1999}	\\
nugent-sn2n	&	SN IIn	&	\citet{Gilliland_high-redshift_1999}	\\
s11-2004hx	&	SN IIL/P	&	\citet{sako_photometric_2011}	\\
s11-2005lc	&	SN IIP	&	\citet{sako_photometric_2011}	\\
s11-2005hl	&	SN Ib	&	\citet{sako_photometric_2011}	\\
s11-2005hm	&	SN Ib	&	\citet{sako_photometric_2011}	\\
s11-2005gi	&	SN IIP	&	\citet{sako_photometric_2011}	\\
s11-2006fo	&	SN Ic	&	\citet{sako_photometric_2011}	\\
s11-2006jo	&	SN Ib	&	\citet{sako_photometric_2011}	\\
s11-2006jl	&	SN IIP	&	\citet{sako_photometric_2011}	\\
hsiao	&	SN Ia	&	\citet{hsiao_k_2007}	\\
hsiao-subsampled	&	SN Ia	&	\citet{hsiao_k_2007}	\\
salt2	&	SN Ia	&	\citet{guy_salt2:_2007}	\\
salt2	&	SN Ia	&	\citet{betoule_improved_2014}	\\
salt2-extended	&	SN Ia	& \citet{pierel_extending_2018}		\\
snf-2011fe	&	SN Ia	&	\citet{pereira_spectrophotometric_2013}	\\
snana-2004fe	&	SN Ic	&	\citet{kessler_results_2010}\\
snana-2004gq	&	SN Ic	&	\citet{kessler_results_2010}	\\
snana-sdss004012	&	SN Ic	&	\citet{kessler_results_2010}	\\
snana-2006fo	&	SN Ic	&	\citet{kessler_results_2010}	\\
snana-sdss014475	&	SN Ic	&	\citet{kessler_results_2010}	\\
snana-2006lc	&	SN Ic	&	\citet{kessler_results_2010}	\\
snana-2007ms	&	SN II-pec	&		\citet{kessler_results_2010}\\
snana-04d1la	&	SN Ic	&	\citet{kessler_results_2010}	\\
snana-04d4jv	&	SN Ic	&	\citet{kessler_results_2010}	\\
snana-2004gv	&	SN Ib	&	\citet{kessler_results_2010}	\\
snana-2006ep	&	SN Ib	&\citet{kessler_results_2010}		\\
snana-2007y	&	SN Ib	&	\citet{kessler_results_2010}	\\
snana-2004ib	&	SN Ib	&	\citet{kessler_results_2010}	\\
snana-2005hm	&	SN Ib	&	\citet{kessler_results_2010}	\\
snana-2006jo	&	SN Ib	&\citet{kessler_results_2010}		\\
snana-2007nc	&	SN Ib	&	\citet{kessler_results_2010}	\\
snana-2004hx	&	SN IIP	&	\citet{kessler_results_2010}	\\
snana-2005gi	&	SN IIP	&	\citet{kessler_results_2010}	\\
snana-2006gq	&	SN IIP	&\citet{kessler_results_2010}		\\
snana-2006kn	&	SN IIP	&\citet{kessler_results_2010}		\\
snana-2006jl	&	SN IIP	&\citet{kessler_results_2010}		\\
snana-2006iw	&	SN IIP	&	\citet{kessler_results_2010}	\\
snana-2006kv	&	SN IIP	&	\citet{kessler_results_2010}	\\
snana-2006ns	&	SN IIP	&\citet{kessler_results_2010}		\\
snana-2007iz	&	SN IIP	&	\citet{kessler_results_2010}	\\
snana-2007nr	&	SN IIP	&\citet{kessler_results_2010}		\\
snana-2007kw	&	SN IIP	&\citet{kessler_results_2010}		\\
snana-2007ky	&	SN IIP	&\citet{kessler_results_2010}		\\
snana-2007lj	&	SN IIP	&\citet{kessler_results_2010}		\\
snana-2007lb	&	SN IIP	&\citet{kessler_results_2010}		\\
snana-2007ll	&	SN IIP	&\citet{kessler_results_2010}		\\
snana-2007nw	&	SN IIP	&\citet{kessler_results_2010}		\\
snana-2007ld	&	SN IIP	&	\citet{kessler_results_2010}	\\
snana-2007md	&	SN IIP	&\citet{kessler_results_2010}		\\
snana-2007lz	&	SN IIP	&	\citet{kessler_results_2010}	\\
snana-2007lx	&	SN IIP	&	\citet{kessler_results_2010}	\\
snana-2007og	&	SN IIP	&	\citet{kessler_results_2010}	\\
snana-2007ny	&	SN IIP	&	\citet{kessler_results_2010}	\\
snana-2007nv	&	SN IIP	&	\citet{kessler_results_2010}	\\
snana-2007pg	&	SN IIP	&	\citet{kessler_results_2010}	\\
snana-2006ez	&	SN IIn	&	\citet{kessler_results_2010}	\\
snana-2006ix	&	SN IIn	&	\citet{kessler_results_2010}	\\
\caption{The SED models used in {\fontfamily{qcr}\selectfont{SNANA}} to describe SN evolution with wavelength and time.}
\label{Atab:snana_seds}
\end{longtable}

\end{document}